 \documentclass[prb,preprint,a4paper,showpacs]{revtex4-1}
 \usepackage{graphicx,amsmath,amssymb}
\newcommand\epjc[3]  {{Eur.\ Phys.\ J. }{\bf C #1} (#2) #3}

\newcommand\plb[3]   {{Phys.\ Lett.\ }{\bf B #1} (#2) #3}
\newcommand\joupr[3]    {{Phys.\ Rev.\ }{\bf #1} (#2) #3}

\newcommand\jouprb[3]   {{Phys.\ Rev.\ }{\bf B #1} (#2) #3}

\newcommand\jouprd[3]   {{Phys.\ Rev.\ }{\bf D #1} (#2) #3}

\newcommand\prep[3]  {{Phys.\ Rept.\ }{\bf #1} (#2) #3}

\newcommand\jourmp[3]   {{Rev.\ Mod.\ Phys.\ }{\bf #1} (#2) #3}

\newcommand\ibid[3]{{ibid.\ }{\bf #1} (#2) #3}

\newcommand{\hepph}[1]{{hep-ph/#1}}

\newcommand{\mathph}[1]{{math-ph/#1}}

\begin{document}

 \title{Consequences of current conservation in systems with partial 
magnetic order}

 \author{K.~Odagiri and T.~Yanagisawa}

 \affiliation{
  Electronics and Photonics Research Institute,
  National Institute of Advanced Industrial Science and Technology,
  Tsukuba Central 2,
  1--1--1 Umezono, Tsukuba, Ibaraki 305--8568, Japan
  }

 \begin{abstract}
  We discuss the consequences of spin current conservation in systems 
with SU(2) spin symmetry that is spontaneously broken by partial 
magnetic order, using a momentum-space approach.
  The long-distance interaction is mediated by Goldstone magnons, whose 
interaction is expressed in terms of the electron Green's functions.
  There is also a Higgs mode, whose excitation energy can be calculated.
  The case of fast magnons obeying linear dispersion relation in three 
spatial dimensions admits nonperturbative treatment using the Gribov 
equation, and the solution exhibits singular behaviour which has an 
interpretation as a tower of spin-1 electronic excitations.
  This occurs near the Mott insulator state.
  The electrons are more free in the case of slow magnons, where the 
perturbative corrections are less singular at the thresholds.
  We then turn our attention to the problem of high-$T_C$ 
superconductivity, through the discussion of the stability of the 
antiferromagnetic ground state in two spatial dimensions.
  We argue that this is caused by an effective mixing of the Goldstone 
$\phi$ and Higgs $h$ modes, which in turn is caused by an effective 
$\phi$ condensation.
  The instability of the antiferromagnetic system is analyzed by 
studying the non-perturbative behaviour of the Higgs boson self-energy 
using the Dyson--Schwinger equations.
 \end{abstract}


 \date{8 April 2011}

 \maketitle

 \tableofcontents

 \section{Introduction}

 \subsection{General picture and framework}

  In this paper, we discuss a momentum-space, Green's function, 
formulation of partial magnetic ordering.
  Our study is motivated phenomenologically by the problem of high-$T_C$ 
superconductivity \cite{highTc}, where the superconducting phase appears 
in between anti-ferromagnetism (AF) and the metallic phase.

  We ordinarily address magnetism (other than through phenomenological 
models) in terms of local models (e.g., the Heisenberg model) or 
Hamiltonians (e.g., the Hubbard model), which start from the picture 
that the electrons are almost localized at individual lattice sites. 
Their interaction, e.g., the $U$ term in the Hubbard model, is written 
under this assumption.

  One fundamental problem which comes with this conventional approach is 
that when the magnetic order is weak and the electrons are relatively 
free, the assumption of locality is no longer viable. Specifically, the 
$U$ term of the Hubbard model is written in a spin-dependent form in 
accord with the Pauli exclusion principle imposed on local 
wavefunctions. This is no longer the case when the wavefunctions are 
non-local.

  One would then ask, how would one formulate the origin of magnetic 
order, which is something that arises at the spatial scale of 
neighbouring lattice sites, in terms of something as delocalized as 
Bloch waves. The answer is that we don't. Instead of having a full 
description of the short-range effects, we parametrize them in terms of 
the energy difference $\Delta E$ between the spin states, and one 
magnetic condensate:
 \begin{equation}
  \left<\psi^*\psi\right>_M\equiv
  \left<\psi^*_g\psi_g\right>-\left<\psi^*_*\psi_*\right>,
  \label{eqn:condensate}
 \end{equation}
  i.e., the density of occupation of electrons with magnetically 
favourable spin, minus the density of occupation of electrons with 
magnetically unfavourable spin.
  $\left<\psi^*\psi\right>_M$ is obviously a function of $\Delta E$, and 
is, at least in principle, calculable for a given dispersion relation 
and total electron density.
  In later discussions, we shall omit the subscript $g$ (which stands 
for `ground') and retain the subscript $*$ (which stands for `excited'). 
Thus
 \begin{equation}
  \left<\psi^*\psi\right>_M\equiv\rho-\rho_*\equiv2S/a^d.
  \label{eqn:spindensity}
 \end{equation}
  $S$ is the average spin per unit cell and $a^d$ is the volume of a 
unit cell. For AF, $S$ should be understood as referring to staggered 
spin.

  The interaction that is responsible for magnetic order is a 
short-distance and short-time effect, and is generally irrelevant when 
discussing long range effects.
  All that we require to know is how the condensate enters into the 
quantities that are relevant to the description of long-distance 
effects, in particular, the Goldstone bosons which correspond physically 
to the magnons\cite{leutwyler}.

  The method by which we shall discuss the Goldstone boson interaction 
is through the electron Green's functions since, on one hand, $\rho$ and 
$\rho_*$ are related to the integrals of the electron Green's functions 
and, on the other hand, the Green's functions are 
constrained\cite{gribov,gribovewsb} by Ward--Takahashi-like 
identities (which, in the following, shall loosely be called the 
Ward--Takahashi identities) through the requirement of spin current 
conservation.

  There is also a Higgs mode, because the high-energy (high-temperature) 
restoration of spin symmetry implies the presence of a mode which 
becomes degenerate with the Goldstone bosons in the limit of zero 
magnetic order.

 \subsection{The case of fast magnons in three spatial dimensions}

  The case of fast (large spin-wave velocity $u$) magnons with linear 
dispersion relation (i.e., like anti-ferromagnetism) in three spatial 
dimensions is of special interest as a model case to illustrate our 
approach.

  When these conditions hold, we can write down a coupled system of 
Gribov equations.
  A Gribov equation \cite{gribov,gribovlectures,yurireview} is a 
non-perturbative self-consistency equation for the electron Green's 
function, which sums the leading overlapping divergences (which plague 
perturbation theory) to all orders.

  We then solve the Gribov equations. The solution exhibits singular 
behaviour of the form $(E-\varepsilon)^{-3/2,1/2}$. We interpret this as 
an indication of the bosonic nature of the electronic excitations. There 
is, furthermore, an infinite tower, with equal separation, of such 
excitations. This is a hint of the presence of a geometrical string (as 
opposed to the string-theory string which is a dynamical object with 
internal vibrational excitations) of variable lengths.

  We do not know how to handle the more general case of magnons with 
arbitrary velocity, but the perturbative correction to the electron 
Green's function is not singular at the thresholds in the limiting case 
of infinitesimally small spin-wave velocity.
  Thus there is some transition from correlated to uncorrelated 
behaviour, depending on the value of $u/v_e$, where $u$ is the magnon 
velocity and $v_e$ is some typical value ($\approx v_F$) of the electron 
velocity.
  The limit of fast magnons corresponds to the case where the vacuum 
responds collectively, and the electronic degrees of freedom is more 
aptly described as being due to string-like objects, whereas the limit 
of slow magnons corresponds to the case where the vacuum does not 
respond fast enough to cause significant alteration to the behaviour of 
individual electrons.

 \subsection{On high-$T_C$ superconductivity}

  Let us turn our attention to the case of two spatial dimensions, still 
with linear dispersion relation for the magnons.

  Since the infrared divergence is now stronger than in the case of 
three spatial dimensions, we expect the long-distance magnetic 
correlation effect to be quite radical and dominant when $u$ is large.
  Superconductivity, in the form of electron pair condensation, thus 
requires $u$ to be small.
  We demonstrate that this is usually the case, unless magnetization $S$ 
is large, $v_e$ is small and the carrier density $g(\mu)$ is small, 
i.e., the system is near the Mott insulator state.

  However, the simple exchange of magnons between electrons is unlikely 
to lead to pair formation.
  Specifically, magnon exchange is attractive between electrons of 
opposite spin.
  Thus spin singlet pairs can form, if the pair binding energy is 
greater than the energy that is required to flip the spin of one of the 
electrons.
  This spin flip goes against the magnetic order which is responsible 
for the appearance of the Goldstone magnons which mediates the 
attractive interaction, and hence pair formation is unlikely.
  In short, we expect that the pair binding energy never exceeds the 
spin flip energy.

  On the other hand, this restriction may be lifted for the case of an 
electron which is borrowed from a neighbouring site.
  As an illustration, the excitation of an $e_*$ electron with an 
up-type spin, let us say, ostensibly costs energy, but one may borrow an 
$e$ electron with an up-type spin from a site where up-type spin is 
favourable, i.e., a site belonging to a different sub-lattice, at a 
lower cost in energy.

  Upon some reflection, we notice that such an effect is an instance of 
Goldstone-boson condensation.
  This then causes an effective Higgs--Goldstone interaction which, in 
turn, makes the AF ground state unstable against the formation of a 
superconducting condensate.
  In view of this, we propose that the analysis should proceed through 
the computation of the Higgs boson Green's function to all orders.
  The instability should then appear in the form of an imaginary part of 
the Higgs boson Green's function.

  The behaviour of the Higgs boson Green's function is then analyzed by 
summing the finite short-distance contribution to the self-energy to all 
orders using a Dyson--Schwinger formalism.
  This leads to an ordinary differential equation governing the 
behaviour of the self-energy as a function of the relevant 
Higgs--Goldstone three-point coupling.
  We solve this equation using a power expansion and Borel summation.

  Some words of caution are necessary here.

  First, an imaginary part of a Green's function is ordinarily 
interpreted as leading to the decay of the particle. This is not the 
case here since there are no final state particles to which the Higgs 
boson can decay. The Higgs boson ostensibly decays into a collective 
state of Goldstone--Higgs mixture, or, the vacuum itself decays.

  Second, the instability of the vacuum does not guarantee that the 
stable new vacuum is superconducting. This statement is quite true in 
general, but we find that the expression for the imaginary part of the 
Green's function turns out to be of the following BCS-like 
\cite{bcs,agd} form:
 \begin{equation}
  \Delta_\mathrm{SC}\sim \omega_\mathrm{cut}
  e^{-1/\mathcal{V N}(\mu)}.
  \label{eqn_electron_magnon_delta_naive}
 \end{equation}
  $\mathcal{V}$ is the 4-point interaction strength. $\mathcal{N}(\mu)$ 
is a suitable density of states.
  Such an exponential form is expected in general in the case of 
pair-wise condensation, and hence we believe, even though this statement 
may not be rigorous, that the ground state is superconducting.

  Third, it may appear strange that the formation of superconducting 
pairs (which have charge $-2e$) may be described by an imaginary part of 
the Higgs boson Green's function, when the Higgs boson is an object 
without electric charge.
  The point here is that the Goldstone--Higgs mixing represents the 
movement of electrons between the sub-lattices, and hence what we are 
summing to all orders is actually the behaviour of the electronic 
condensate. In other words, a Higgs boson is not an exciton-like 
electron--hole object but the fluctuation in the number density of 
electrons.

  Fourth and last, when we discuss short-distance physics, we should in 
general be wary of the effect of short-distance interaction which, as we 
discussed earlier, are hidden and neglected.
  However, this applies primarily to the electrons, and not to the 
Goldstone and Higgs bosons which are collective excitations.
  Furthermore, the short-distance contributions which we are summing to 
all orders are not necessarily as short as the neighbouring lattice 
sites.

  The calculated mean-magnetization dependence of the energy gap is 
consistent with the observed doping dependence of the superconducting 
critical temperature $T_C$ in high-$T_C$ compounds.
  Our analysis suggests the co-existence of partial AF order with 
superconductivity in the under-doped region.

 \subsection{Organization of the paper}

  This paper is organized as follows.

  We present a general discussion of spin-current conservation and the 
bosonic modes in systems with partial magnetic order in 
sec.~\ref{sec_spin_current}.

  We derive and solve the Gribov equation for electrons interacting with 
fast Goldstone magnons in three spatial dimensions in 
sec.~\ref{sec_griboveqn}.

  We discuss the case of two spatial dimensions and high-$T_C$ 
superconductivity in sec.~\ref{sec_highTc}.

  The conclusions are stated at the end.

 \section{General analysis of spin current conservation}
 \label{sec_spin_current}

  In this section, we adapt Gribov's analysis of axial current 
conservation \cite{gribov,gribovewsb} to the context of spin current 
conservation in systems with partial magnetic order.

  Consider the fermionic (quasi-)electron field which is an 
SU(2)$_\mathrm{spin}$ doublet
 \begin{equation}
  \Psi=\left(\begin{array}{c}\psi_\uparrow\\\psi_\downarrow
  \end{array}\right).
 \end{equation}
  Let us define the four-vector currents. For $\widehat\Gamma^\mu$ 
defined by 
the conserved electron current
 \begin{equation}
  J^\mu_\mathrm{electron}=\Psi^\dagger\widehat\Gamma^\mu\Psi,
  \label{eqn_electron_current}
 \end{equation}
  we define $4$-vector spin currents $J^\mu_i$ ($\mu=0\ldots3$, 
$i=1,2,3$) of the form
 \begin{equation}
  J^\mu_i=\Psi_1^\dagger\sigma_i\widehat\Gamma^\mu\Psi_2.
  \label{eqn_spin_current}
 \end{equation}
  For example, $J^\mu_3$ is the up-spin current minus down-spin current.
  $\widehat\Gamma^\mu$ can be an operator, e.g., 
$\widehat\Gamma^\mu=i\partial/\partial x_\mu$.

  Let us now move to the momentum space. The current vertex $\Gamma^\mu$ 
is the momentum-space counterpart of $\widehat\Gamma^\mu$ in the real 
space.
  In the spin-symmetric phase, the spin currents are conserved because 
of the conservation of electrons.
  That is, the following Ward--Takahashi identity is satisfied:
 \begin{equation}
  \Gamma^\mu (q_1-q_2)_\mu\propto 
G^{-1}(q_1,\lambda_1)-G^{-1}(q_2,\lambda_2).
  \label{eqn_ward_takahashi}
 \end{equation}
  $q$ are $d+1$-vectors with components $(q_0,\mathbf{q})$.
  $q_0$ is the energy. $\mathbf{q}$ is either the spatial momentum or 
the wave number (we shall be sloppy later on for the sake of the brevity 
of notation).
  $\lambda_{1,2}$ refer to the spin states, but these are dummy indices 
here in the sense that $G^{-1}(q,\lambda)$ is independent of $\lambda$.

  The Ward--Takahashi identity is violated in the symmetry-broken 
phase, since there is now an energy difference between the different 
spin states.

  We can set locally, and without loss of generality,
 \begin{equation}
  \Psi=\left(\begin{array}{c}\psi_*\\\psi\end{array}\right).
 \end{equation}
  $\psi$ is the spin ground state and $\psi_*$ is the excited state.
  Even in the case of AF, we can still denote the spin ground state and 
the excited state as $\psi$ and $\psi_*$, respectively.

  Let us define the energy difference $\Delta E$ between the two spin 
states by
 \begin{equation}
  \Delta E= G^{-1}(q)-G^{-1}_*(q).
  \label{eqn_delta_E_definition}
 \end{equation}
  If $G^{-1}$ and $G^{-1}_*$ are both linear in energy, $\Delta E$ is 
given by $\epsilon_*-\epsilon$ and is constant up to a possible 
dependence on the spatial momentum $\mathbf{q}$.
  $\Delta E$ depends on the energy $q_0$ in principle.
  In particular, at the threshold, $G^{-1}$ would, in general, have a 
singular structure corresponding to the emission and absorption of the 
Goldstone boson $\phi$ through the process $e_*\to e\phi$.
  We shall encounter a case in sec.~\ref{sec_griboveqn} where the 
electron Green's function is strongly singular in the threshold regions. 
  The electrons move in a correlated fashion, and $\Delta E$ cannot be 
treated as a constant under such circumstances.
  However, $\Delta E\approx\mbox{const.}$ is a reasonable approximation 
when 
the correlation is weak and the decay width for $e_*\to e\phi$ is 
relatively small.

  After the symmetry violation, the currents $J^\mu_{1,2}$ are no longer 
conserved, and the Ward--Takahashi identity is violated by
 \begin{equation}
  \Gamma^\mu (q_1-q_2)_\mu\propto 
G^{-1}(q_1,\lambda_1)-G^{-1}_2(q_2,\lambda_2)
  \pm\Delta E\quad (\lambda_1\ne\lambda_2).
  \label{eqn_ward_takahashi_violation}
 \end{equation}
  This $\Delta E$ contribution is of the same form as the coupling of 
the Goldstone boson $\phi$, and current conservation is restored by 
including the contribution of the Goldstone boson. This is the case even 
when $\Delta E$ is not constant.
  Specifically, spin current conservation is restored for the vertex 
modified by the inclusion of the Goldstone boson,
 \begin{equation}
  \includegraphics[width=6.5cm]{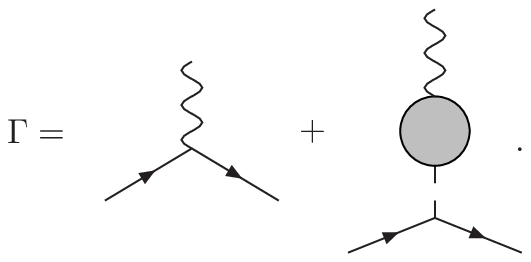}
  \label{eqn_modified_vertex}
 \end{equation}
  The wavy lines indicate the spin current, and the dashed line 
indicates the Goldstone boson.
  The two-point function in the second term is nothing but an electron 
loop.
  There is nothing strange in this result, since the Goldstone boson 
arose in the first place as the longitudinal component of the spin 
current.
  After taking away the longitudinal component, the remaining part is 
transverse and therefore satisfies the Ward--Takahashi identity.

  The Goldstone bosons $\phi_1$ and $\phi_2$ correspond to the SU(2) 
rotation,
 \begin{equation}
  \mathcal{U}(\phi_1,\phi_2)=\exp
  \left[if^{-1}\sum_{i=1}^{2}\phi_i\sigma_i\right],
 \end{equation}
  and they correspond physically to the magnons.
  $f$ is the Goldstone boson form factor which, by virtue of 
eqn.~(\ref{eqn_modified_vertex}), is calculated as the strength of the 
current--Goldstone-boson two-point amplitude.
  Their coupling strength with fermions is then given by $f^{-1}\Delta 
E$.
  The Higgs boson mass $M_h$, or excitation energy, is evaluated in a 
similar fashion \cite{gribovewsb} to the evaluation of $f$.

 \subsection{Case of quadratic dispersion relation}
 \label{sec_quadratic_dispersion}

  For ferromagnetism, the spin-wave velocity is known to have a 
quadratic dispersion relation, $\omega\propto \mathbf{k}^2$, for small 
$\mathbf{k}$. That is, the Green's function is given by 
$D^{-1}_\phi=E-c\mathbf{k}^2$, where $c$ is some constant.
  In this case, evaluating the electron loop in the two-point function 
shown in the second term of eqn.~(\ref{eqn_modified_vertex}), between 
the spin current and the Goldstone boson, should yield
$f\times(1,c\mathbf{q})$ in the soft limit such that 
eqn.~(\ref{eqn_modified_vertex}) satisfies the Ward--Takahashi identity.
  The form factor $f$ has the dimension of (length)$^{-d/2}$.

  By considering the 0th component of the two-point function in the soft 
limit, we obtain
 \begin{equation}
  f^2=-2\times\int\frac{d^{d+1}k}{(2\pi)^{d+1}i}
  G(k)G_*(k)\Delta E(k).
  \label{eqn_ferromagnetic_form_factor_as_loop_integral}
 \end{equation}
  The factor $2$ is from contracting the Pauli matrices, or for the 
two spin orientations.
  The corrections to the electron Green's function are included in this 
expression.
  As for the vertex corrections, we are taking the current vertex to be 
bare and take the value $1$ in its 0th component, whereas the 
Goldstone-boson vertex is renormalized (as otherwise there will be 
double counting).
  The latter correction gives rise to the renormalization of $\Delta E$. 
$f$ is just a number.
  This expression, and the results that follow, are therefore exact.

  By eqn.~(\ref{eqn_delta_E_definition}), and because the integral of a 
Green's function is the number density, we find
 \begin{equation}
  f^2=2(\rho-\rho_*).
 \end{equation}
  $\rho$ is the number per unit volume or area.

  We can define the mean magnetization $S$, $0\le S\le1/2$, by
 \begin{equation}
  2S/a^d=\rho-\rho_*,
 \end{equation}
  to obtain
 \begin{equation}
  f^2=4S/a^d.
  \label{eqn_fsq_S_ferromagnetism}
 \end{equation}

  Let us temporarily consider a magnetic order-parameter field $\Phi$, 
whose rotational invariances correspond to the Goldstone bosons.
  It is easy to see that $f=2v$, where $v$ is the vacuum expectation 
value of $\Phi$.
  This implies $v^2=S/a^d$, which seems natural.

 \subsection{Case of linear dispersion relation}
 \label{sec_linear_dispersion}

  In the case of a linear dispersion relation for the Goldstone boson as 
in the AF magnons, let us proceed analogously to the above case of 
quadratic dispersion relation.
  Now the Green's function of the Goldstone boson is given by 
$D^{-1}_\phi=E^2-\hbar^2u^2\mathbf{k}^2$.
  We then see that the two-point function shown in the second term of 
eqn.~(\ref{eqn_modified_vertex}), between the spin current and the 
Goldstone boson, is now given by $fq_\mu$ (with the spin-wave velocity 
$u=1$), where $q_\mu$ is the $4$-momentum.
  In particular, the two-point function vanishes in the soft limit. In 
the case of AF, this is due to the cancellation between the two spin 
orientations, or between the two sub-lattices.

  The form factor $f^{-1}$ has the dimension of 
(energy$\times$length$^d$)$^{1/2}$. Let us make use of
 \begin{equation}
  \includegraphics[width=8cm]{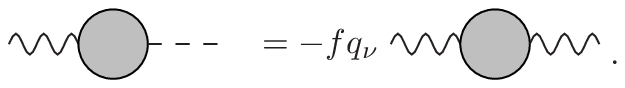}
 \end{equation}
  Note that we have been rather loose in the notation here: this 
equation is valid only in the sense that the cancellation between the 
sub-lattices has been taken into account on the left-hand side.
  If not, we would be equating something that is finite in the soft 
limit on the left-hand side against something which vanishes on the 
right-hand side.

  It follows that, for small momenta,
 \begin{equation}
  \includegraphics[width=5.5cm]{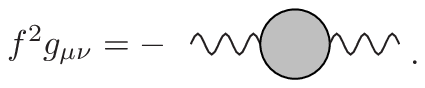}
  \label{eqn_form_factor_final}
 \end{equation}
  For the 0--0 component of this quantity, and for nearly constant 
$\Delta E$, we obtain
 \begin{equation}
  f^2=2\times\int\frac{d^d\mathbf{k}}{(2\pi)^d}\frac1{\Delta E}.
 \end{equation}
  The factor $2$ is again from contracting Pauli matrices. The region of 
integration covers all points where one spin state is occupied and the 
other spin state is vacant.

  Let us define the mean magnetization $S$, $0\le S\le1/2$, again by
 \begin{equation}
  2S/a^d=\rho-\rho_*,
 \end{equation}
  We then have
 \begin{equation}
  f^2=\frac{4S}{a^d\Delta E}.
  \label{eqn_fsq_S}
 \end{equation}
  In terms of the order-parameter description, $f=2v$ as before, and 
this expression seems natural from this perspective.

 \subsection{Relationships among $\Delta E$, 
$\left<\psi^*\psi\right>_M$, $S$ and $u$}
 \label{subsec_delta_e_S}

  $\left<\psi^*\psi\right>_M$ is controlled by $\Delta E$ when $\Delta 
E$ is nearly constant. If we denote the density of states (of both spin 
states) by $g(E)$ and if this density of states can be taken to be 
roughly constant near the chemical potential $\mu$, we see immediately 
by comparing the occupation of the two spin states that
 \begin{equation}
  \rho-\rho_*=g(\mu)\Delta E/2,
 \end{equation}
  or
 \begin{equation}
  \Delta E=\frac{4S}{g(\mu)a^d}.
  \label{eq:delta_e_S}
 \end{equation}
  Hence
 \begin{equation}
  f^2=\left\{\begin{array}{l}
  g(\mu)\Delta E\quad \mbox{(quadratic dispersion relation)},\\
  g(\mu)\quad \mbox{(linear dispersion relation)}.
  \end{array}\right.
  \label{eq:fsq_gmu}
 \end{equation}

  In secs.~\ref{sec_quadratic_dispersion} and 
\ref{sec_linear_dispersion}, we calculated $f^2$ through the 
consideration of the time component of the spin-current--magnon 
two-point function.
  Looking at the space component of the same amplitude should give us 
the spin-wave velocity, in principle.
  However, the calculation is messy, and it can depend on the threshold 
behaviour of $G$, $G_*$ and $\Delta E$.

  A simpler and more intuitive way to obtain the spin-wave velocity, for 
a given form of the dispersion relation, is to equate the excitation 
energy $\hbar\omega_\phi$ of the $\pi/a$ magnon mode with the excitation 
energy of $e$ to $e_*$ at $\pi/a$, because these are equivalent. This 
tells us, quite generally, that
 \begin{equation}
  \hbar\omega_\phi(\pi/a,0,0)=
  \varepsilon_*(\pi/a,0,0)-\varepsilon(\pi/a,0,0)=\Delta E.
 \end{equation}
  In particular, for a linear dispersion relation with velocity $u$, we 
obtain
 \begin{equation}
  \frac{\hbar u\pi}a=\Delta E \qquad\Rightarrow\quad
  u=\frac{a\Delta E}{\hbar\pi}=\frac{4S}{\hbar\pi g(\mu)a^{d-1}}.
  \label{eq_linear_spin_wave_velocity}
 \end{equation}
  Note that we cannot expect the dispersion relation to be exactly 
linear up to $\pi/a$.
  This is an approximation.

 \subsection{The Higgs boson}
 \label{subsec_higgs_boson}

  As we expect spin symmetry to be restored in the high energy 
(temperature) limit, there is necessarily a Higgs boson $h$ to 
complement the Goldstone bosons and provide the remaining local 
rotational degree of freedom.
  By this symmetry, the Higgs boson couples to fermions with the same 
coupling strength $(f^{-1}\Delta E)$ as the Goldstone bosons, but there 
is now an energy gap.
  This energy gap is due to corrections below the high energy scale.
  Explicitly, the corrections are given by the self-energy integral:
 \begin{equation}
  \Sigma_h=-\int\frac{d^{d+1}k}{(2\pi)^{d+1}i}
  (f^{-1}\Delta E)^2 
  \left[G(k)G(k-q)+G_*(k)G_*(k-q)\right].
  \label{eq:higgs_self_energy}
 \end{equation}
  We then compare this with the corresponding integral for the Goldstone 
bosons:
 \begin{equation}
  \Sigma_\phi=-2\int\frac{d^{d+1}k}{(2\pi)^{d+1}i}
  (f^{-1}\Delta E)^2 G(k)G_*(k-q).
 \end{equation}
  We see that
 \begin{equation}
  \Sigma_h\left(q+(\Delta E,\mathbf{0})\right)=\Sigma_\phi(q),
 \end{equation}
  i.e., the correction to the Higgs boson at energy $\Delta E+E$ is the 
same as that to the Goldstone boson at energy $E$.
  Since the Goldstone bosons have zero excitation energy, the excitation 
energy for the Higgs boson is thus equal to $\Delta E$ (at least for 
almost constant $\Delta E$).
  It follows that the Green's functions for the Higgs boson are given 
by
 \begin{equation}
  D^{-1}_h(q)=D^{-1}_\phi(q)-\left\{\begin{array}{l}
  \Delta E\quad \mbox{(quadratic dispersion relation)},\\
  (\Delta E)^2\quad \mbox{(linear dispersion relation)}.
  \end{array}\right.
  \label{eq:higgs_gf}
 \end{equation}

  The above derivation is somewhat intuitive. Let us derive the same 
results more rigorously, using a UV counterterm argument.
  The excitation energy is ordinarily found by the condition that the 
radiative correction vanishes, i.e., $\Sigma=0$, when the excitation is 
on the mass shell.
  This $\Sigma$ should be the sum of $\Sigma_h$ and $\Sigma_{UV}$, where 
$\Sigma_{UV}$ is the high-energy contribution due to 
$\left<\psi^*\psi\right>_M$, and is the same between $\phi$ and $h$.
  The condition that $\Sigma_\phi+\Sigma_{UV}=0$ at $q=0$ completely 
fixes $\Sigma_{UV}$ as
 \begin{equation}
  \Sigma_{UV}=+2\int\frac{d^{d+1}k}{(2\pi)^{d+1}i}
  (f^{-1}\Delta E)^2 G(k)G_*(k)=-2f^{-2}\Delta 
  E\left<\psi^*\psi\right>_M.
 \end{equation}
  From the requirement $\Sigma_h+\Sigma_{UV}=0$, we then obtain the 
excitation energy $\Delta E$. That there is a solution to $\Sigma=0$ is 
a sufficient proof for the existence of the Higgs boson.

 \section{The non-relativistic Gribov equation}
 \label{sec_griboveqn}

 \begin{figure}[ht]{
  \centerline{
   \includegraphics[width=5cm]{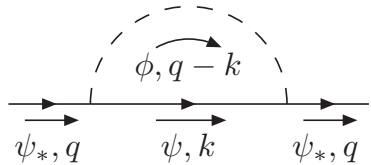}
  }
  \caption{The electron self-energy diagram.
  \label{fig_one_loop_electron_self_energy}}}
 \end{figure}

  The exchange of Goldstone bosons is the leading long-distance 
contribution to the electron self-energy.
  Corresponding to fig.~\ref{fig_one_loop_electron_self_energy}, the 
self-energy is given by
 \begin{equation}
  \Sigma_*(q)=2\int\frac{d^{d+1}k}{(2\pi)^{d+1}i}
  f^{-2}\Delta E^2 G(k)D(q-k),
  \label{eq:electron_self_energy_general}
 \end{equation}
  and similarly for $\Sigma(q)$.
  The factor 2 sums over the two Goldstone modes.
  $D$ is the propagator of the Goldstone bosons. This expression 
contains all corrections other than the vertex correction which leads to 
a correction to $\Delta E$.

  In general, $G(k)$ and $G_*(k)$ become singular in the threshold 
regions where the decays $e_*\to e\phi$ and $e\to e_*\phi$ (the latter 
being to a virtual $e_*$) open up.
  This is fine at the leading order, but since corrections to $G(k)$ and 
$G_*(k)$ affect $\Delta E$, the general expression becomes troublesome, 
and we do not know how to handle it. This problem of `overlapping 
divergences' is, of course, a well-known generic problem of perturbation 
theory.

  However, we have one exception in the case of three spatial 
dimensions, with a linear dispersion relation of the Goldstone boson, 
and with large $u$. We may, in this case, employ the Gribov equation 
framework \cite{gribov,gribovlectures,yurireview}. Let us therefore 
specialize to this situation here.

 \subsection{Derivation of the Gribov equation}

  To obtain the Gribov equation, we apply
 \begin{equation}
  \frac{\partial^2}{\partial q_0^2}-
  \frac1{u^2}\frac{\partial^2}{\partial\mathbf{q}^2},
 \end{equation}
  to eqn.~(\ref{eq:electron_self_energy_general}).
  The integral sign then disappears because of the following identity:
 \begin{equation}
  \left[\frac{\partial^2}{\partial q_0^2}-
  \frac1{u^2}\frac{\partial^2}{\partial\mathbf{q}^2}\right]
  \frac1{(q_0-k_0)^2-u^2(\mathbf{q}-\mathbf{k})^2+i\epsilon}=
  \frac{4\pi^2i}{u^3}\delta^{(4)}(q-k).
  \label{eqn_delta_emergence}
 \end{equation}
  In other words, $D(q)$ is a Green's function of the 4-dimensional 
Laplace equation in the energy--momentum space.
  We have assumed that the boundaries in $\mathbf{k}$ do not affect the 
argument, because $u$ is large. That is, only small $\mathbf{k}$ is 
exchanged, typically.
  We then obtain
 \begin{equation}
  \left(\frac{\partial^2}{\partial q_0^2}-
  \frac1{u^2}\frac{\partial^2}{\partial\mathbf{q}^2}\right)
  \Sigma_*(q)
  =\frac{(f^{-1}\Delta E)^2}{2\pi^2\hbar^3u^3} G(q).
  \label{eqn_gribov_oneloop}
 \end{equation}

  So far, this is just an algebraic manipulation which is valid at the 
one-loop order, but we now notice that something else has occurred to the 
right-hand side.
  To see this, let us consider the general $N$-loop perturbative 
contribution $\delta\Sigma$ to the self-energy.
  The leading contribution is from regions with overlapping divergences, 
where there may be a number of large logarithms being involved, in the 
form, for example,
 \begin{equation}
  \delta \Sigma\sim\log(r_1)\log(r_2),
 \end{equation}
  where $r_1$ and $r_2$ are some large ratios.
  Now we consider differentiating this expression twice. If we 
differentiate $\log(r_1)$ once and $\log(r_2)$ once, we obtain something 
like $1/r_1r_2$, and both of the large logarithms have disappeared.
  On the other hand, if we differentiate only $\log(r_1)$, the other 
logarithms remain intact, and so this is the leading contribution to the 
double derivative of $\delta\Sigma$.

  Since these large logarithms are due to the integrals of $D(q)$, it 
follows that the double derivative, when applied to the general $N$-loop 
expression for self-energy, removes one or the other of the Goldstone 
boson propagators, and, the final contribution is of the form given in 
eqn.~(\ref{eqn_gribov_oneloop}).
  The higher loop contributions amount to the vertex corrections, and so 
eqn.~(\ref{eqn_gribov_oneloop}) is in fact valid to all orders in the 
sense of summing the leading overlapping divergences.

  Note that this approach is quite distinct from methods such as the 
leading-logarithm summation, which do not deal with the problem of 
overlapping divergences. More detailed discussion of the various 
contributions to the Gribov equation is available in the 
papers\cite{gribov} by Gribov.

  Equation (\ref{eqn_gribov_oneloop}) can be turned into a 
self-consistency equation for the electron Green's function in the limit 
of large $u$, since the left-hand side then becomes equivalent to the 
energy double derivative of $-G^{-1}_*(q)$, i.e.,
 \begin{equation}
  -\frac{\partial^2}{\partial q_0^2} G^{-1}_*(q)
  =\frac{(f^{-1}\Delta E)^2}{2\pi^2\hbar^3u^3} G(q).
  \label{eqn_gribov_nr_step_a}
 \end{equation}
  Note that, in general, $G^{-1}=G_0^{-1}-\Sigma$, and we are assuming 
that $G_0^{-1}$ has the form $G_0^{-1}=q_0-\varepsilon(\mathbf{q})$.
  The spatial-momentum dependence has dropped out of the expression, and 
we can write this equivalently as
 \begin{equation}
  \left\{
   \begin{array}{l}
   (G^{-1}_*)''=-f_R^{-2}(G^{-1}-G^{-1}_*)^2 G,\\
   (G^{-1})''  =-f_R^{-2}(G^{-1}-G^{-1}_*)^2 G_*,
   \end{array}
   \right.
  \label{eqn_gribov_nr}
 \end{equation}
  with $f_R^2=2\pi^2\hbar^3u^3f^2$. Prime refers to the energy 
derivative.

 \subsection{Solution of the Gribov equation}
 \label{subsec_gribov_solution}

  Let us proceed to solve eqns.~(\ref{eqn_gribov_nr}).
  We first consider the expression
 \begin{equation}
  G_*(G_*^{-1})''-G(G^{-1})''=0,
 \end{equation}
  which follows from eqns.~(\ref{eqn_gribov_nr}).
  We then use the identity $x^{-1}x''\equiv (\ln x)''+((\ln x)')^2$:
 \begin{equation}
  (\ln G_*^{-1}/G^{-1})''+((\ln G_*^{-1})')^2-((\ln G^{-1})')^2=0.
 \end{equation}
  Let us define $z=G_*^{-1}/G^{-1}$, so that
 \begin{equation}
  (\ln z)''+(\ln z)'(\ln G_*^{-1}G^{-1})'=0.
 \end{equation}
  This can be integrated once, and we obtain
 \begin{equation}
  \ln(\ln z)'+\ln(G_*^{-1}G^{-1})=\mbox{const.},
 \end{equation}
  or,
 \begin{equation}
  (\ln z)'=CGG_*.
 \end{equation}
  $C$ is a constant of integration. This implies
 \begin{equation}
  G^2=C^{-1}z', \quad G_*^2=C^{-1}z'/z^2.
  \label{eq:gribovConstofIntegration}
 \end{equation}

  Let us make use of eqn.~(\ref{eq:gribovConstofIntegration}) to 
eliminate the energy derivatives in eqns.~(\ref{eqn_gribov_nr}). This 
gives us, almost trivially,
 \begin{equation}
  G^3\frac{d^2}{dz^2}G=(f_RC)^{-2}(z+z^{-1}-2).
  \label{eq:gribovGvsz}
 \end{equation}
  The equation for $G_*$ is obtained by substituting $G_*$ for $G$ and 
$z^{-1}$ for $z$.

  This equation is simple, but does not seem to have a simple analytical 
solution.
  However, we obtain the following limiting behaviours:
 \begin{itemize}
  \item When $G$ diverges, $z=G_*^{-1}/G^{-1}$ should also diverge. We 
see that the limiting expressions are given by 
$G=(f_RC)^{-1/2}(-16z^3/3)^{1/4}$ and $G_*=(f_RC)^{-1/2}(-16/3z)^{1/4}$. 
That is, $G_*$ vanishes.
  \item When $G_*$ diverges, $G$ and $z$ vanish. The limiting 
expressions are given by $G=(f_RC)^{-1/2}(-16z/3)^{1/4}$ and 
$G_*=(f_RC)^{-1/2}(-16/3z^3)^{1/4}$.
 \end{itemize}

  In order to obtain a numerical solution of eqn.~(\ref{eq:gribovGvsz}), 
it is desirable to replace $z$ with something that has a limited range.
  In this regard, we notice that the left-hand side of 
eqn.~(\ref{eq:gribovGvsz}) has a conformal symmetry with respect to 
transformations on $z$. This symmetry becomes more manifest when we 
return to eqns.~(\ref{eqn_gribov_nr}) and now eliminate $G$ and $G_*$ 
using eqn.~(\ref{eq:gribovConstofIntegration}).
  After some elementary algebra, we obtain
 \begin{equation}
  \frac{z'''}{2z'}-\frac34\left(\frac{z''}{z'}\right)^2
  =f_R^{-2}(z+z^{-1}-2).
  \label{eq:gribovODE}
 \end{equation}
  We then see that the left-hand side of this equation is invariant 
under the M\"{o}bius transformation:
 \begin{equation}
  z\longrightarrow w=\frac{az+b}{cz+d}.
 \end{equation}
  It turns out that $z$ is negative. Let us therefore use
 \begin{equation}
  w=\frac{1+z}{1-z},\quad z=-\frac{1-w}{1+w}\qquad (-1\le w\le1).
  \label{eq:mobius}
 \end{equation}
  We also define
 \begin{equation}
  y=(16f_R^{-2}/3)^{-1/4}\sqrt{\frac{dw}{dE}}
  =\frac{(3f_R^2C^2/4)^{1/4}}{G^{-1}-G_*^{-1}}
  \equiv\frac{(3f_R^2C^2/4)^{1/4}}{\Delta E}.
  \label{eq:gribov_ydef}
 \end{equation}
  We then obtain
 \begin{equation}
  y^3\frac{d^2y}{dw^2}=-\frac{3}{4(1-w^2)}.
  \label{eq:gribovyvsw}
 \end{equation}
  Note the similarity with eqn.~(\ref{eq:gribovGvsz}). We may have 
obtained eqn.~(\ref{eq:gribovyvsw}) directly from 
eqns.~(\ref{eq:gribovGvsz}) and (\ref{eq:gribovConstofIntegration}) 
using eqn.~(\ref{eq:mobius}).

\begin{figure}[ht]
  \centerline{\includegraphics[width=12cm]{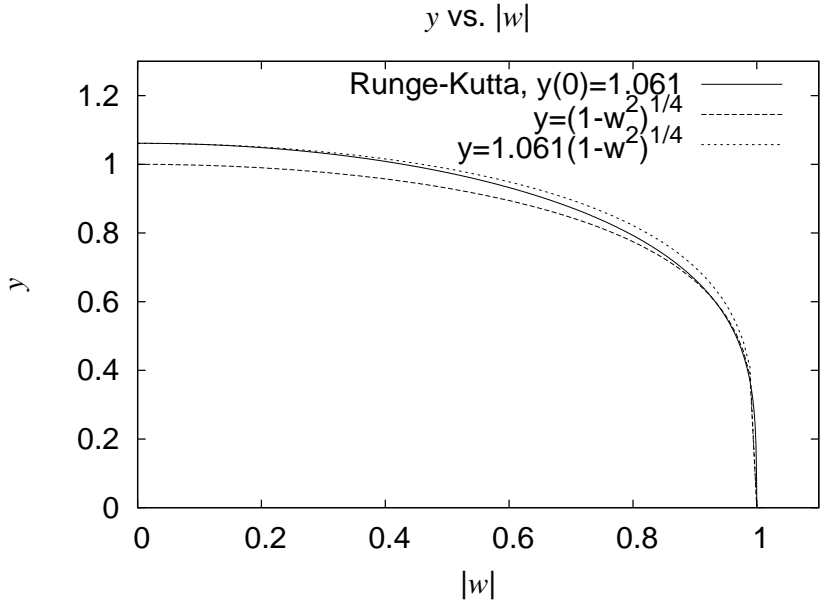}}
  \caption{$y$ vs $w$. We compare the numerical results with the 
limiting expression $y\propto(1-w^2)^{1/4}$. \label{fig_gribovyvsw}}
 \end{figure}

  Now $y$ vanishes when $w=\pm1$, and the limiting expression is given 
by $y\to(1-w^2)^{1/4}$.
  We should use this boundary condition to solve 
eqn.~(\ref{eq:gribovyvsw}) numerically, but this is difficult.
  We therefore use the boundary condition $dy/dw=0$ at $w=0$ and 
substitute some numbers for $y$ at $w=0$ and, by trial and error, 
establish the value of $y$ at $w=0$ that leads to the correct behaviour 
at $w=\pm1$.
  The result of this calculation is shown in fig.~\ref{fig_gribovyvsw}.
  The numbers were obtained using the classical Runge--Kutta method with 
the step size of 0.0005, and were found to be stable against 
modifications to the step size.

  We found that the value of $y$ is approximately $1.061$ when $w=0$.
  We see that $y=(1-w^2)^{1/4}$ and $y=1.061(1-w^2)^{1/4}$ are both good 
approximations to the actual behaviour of $y$.

  Let us define the scaled energy $x$ as 
$x=(E-\varepsilon_*)\sqrt{16f_R^{-2}/3}$, so that $y,z\to0$ as $x\to0$.
  By eqn.~(\ref{eq:gribov_ydef}), we obtain
 \begin{equation}
  \int dx=\int\frac{dw}{y^2(w)}.
 \end{equation}
  The integral on the right-hand side is finite when the integration is 
performed between $w=\pm1$.
  When $w=\pm1$, the derivative of $w$ with respect to $x$ vanishes.
  $w$ is therefore periodic in $x$.
  Let us, for the sake of simplicity, adopt $y=(1-w^2)^{1/4}$. This 
leads to
 \begin{equation}
  w=\cos x,\quad z=-\tan^2(x/2).
  \label{eqn_gribov_nr_approx_solution}
 \end{equation}

 \subsection{Nature of the solution}

  We have seen in the above that the solution has the following form:
 \begin{equation}
  z\approx-\tan^2(x/2)
 \end{equation}
  This implies that, by eqn.~(\ref{eq:gribovConstofIntegration}),
 \begin{equation}
  G\approx\pm\sqrt{-C^{-1}\tan(x/2)\sec^2(x/2)},\quad 
  G_*\approx\mp\sqrt{-C^{-1}\cot(x/2)\mathrm{cosec}^2(x/2)}.
 \end{equation}
  As $x\to0$, these functions behave as
 \begin{equation}
  G\to\pm\sqrt{-C^{-1}x/2},\quad G_*\to\mp\sqrt{-8C^{-1}/x^3}.
  \label{eq:x_to_zero_behaviour}
 \end{equation}
  This is clearly a rather exotic behaviour for the Green's function of 
an electron. For instance, the integral of the Green's function, or the 
electron number, becomes undefinable.
  Note that this does not necessarily invalidate our starting point 
since we can still define $\left<\psi^*\psi\right>_M$ as a finite 
number.
  Furthermore, the non-local correlator $\left<\psi^*(t)\psi(t+\Delta 
t)\right>$ is definable and is non-zero (at, in this case, discrete 
values of $\Delta t$). That is, the field theory is non-local.

  One interpretation of this exotic power behaviour is through the 
Levinson theorem which states that the number of states in between 
energies $E_1$ and $E_2$ is given by the difference in the phase shifts 
$\delta(E_1, E_2)$ under scattering:
 \begin{equation}
  N=\frac1\pi\left[\delta(E_2)-\delta(E_1)\right].
 \end{equation}
  In this sense, we can say that there is $-1/2$ electron of the $e$ 
type and $+3/2$ electron of the $e_*$ type at $x=0$.
  On the other hand, when $z$ diverges, we obtain $-1/2$ $e_*$ electron 
and $+3/2$ $e$ electron.

  This result indicates that there is no longer a physical electron. 
However, if we sum the states that are present at $x=0$, we see that 
there is $-1/2+3/2=1$ state, whose charge is $-e$ and spin is, if such a 
sum may be made, $\left|-1/2(1/2)+3/2(-1/2)\right|=1$.
  In this sense, the electronic excitations are bosonic.

  What is the nature of these bosonic excitations?

  Our solution implies that there is a tower of such states, at each 
phase-space point, with equal separation in energy (of approximately 
$f_R\pi\sqrt3/4$) and alternating spin direction.
  Note that the number of states in each excited state is the same. This 
implies that the excitations are in the form of one-dimensional objects. 
  This then suggests the existence of string-like objects (such as the 
trajectory drawn by spin-flipped electrons), which bind together the 
electrons. The quantized energies correspond to the length of the 
string.
  The string is not a dynamical object with an internal degree of 
freedom, as otherwise we cannot explain the one-dimensional nature of 
the excitations.

  We should add that a power-like behaviour of the Green's function is 
obtained also in the case of simple photon-exchange interaction. In this 
case, the Gribov equation is given by
 \begin{equation}
  (G^{-1})^{\prime\prime}=\frac{\alpha}{\pi}G((G^{-1})^{\prime})^2.
 \end{equation}
  $\alpha\approx1/137$ is the fine structure constant. This is in the 
absence of screening. It is easy to see that this has the following 
general solution:
 \begin{equation}
  G^{-1}\propto (E-\varepsilon)^{1/(1-\alpha/\pi)}.
 \end{equation}
  According to the above argument based on the Levinson theorem, the 
number of electronic states goes up as a result of the repulsive Coulomb 
interaction. There seems to be a contribution due to plasmon-like charge 
oscillation states here.

 \subsection{The case of slow magnons}

  The above discussion is for the case of fast magnons with linear 
dispersion relation in three spatial dimensions.
  This case is special in that the leading overlapping divergences can 
be dealt with by using the Gribov equation framework.

  Let us consider relaxing one or the other of these conditions.

  First, we do not know what the situation may be with respect to 
changing the linear dispersion relation to a quadratic dispersion 
relation.

  Second, the infra-red divergence is stronger in 2-d. We thus expect 
the long-distance correlations to be even more dominant than in the 3-d 
case.
  On the other hand, we know that there is no long-distance correlation 
in the 1-d case because of the strong quantum fluctuations.
  Our intuitive guess (which is only a wild guess) is that the string 
picture is valid both for 3-d and 2-d, because the string is a 
one-dimensional object which can be embedded in both 3-d and 2-d.
  The 1-d case is obviously different in this picture.

  Third and last, let us consider the case of very slow magnons. This is 
interesting because although 
eqn.~(\ref{eq:electron_self_energy_general}) diverges, it is almost 
constant with respect to variations in the external 4-momentum $q$. 
Explicitly, let us consider replacing 
eqn.~(\ref{eq:electron_self_energy_general}) by the expression
 \begin{equation}
  \left.\Sigma_*(q)\right|_{\mbox{small $u$}}\approx
  2\int\frac{d^{d+1}k}{(2\pi)^{d+1}i}
  f^{-2}\Delta E^2 \frac1{(q_0-k_0)^2-u^2\kappa^2+i\delta}
  \frac1{k_0-\mathbf{k}^2/2m\pm i\delta}
  \label{eq:electron_self_energy_slow}
 \end{equation}
  in the limit of small $u$. $\kappa$ is some typical spatial momentum 
scale. Then this expression is obviously independent of $\mathbf{q}$, 
and it can only depend mildly on $q_0$. More explicitly, we find that, 
as we would naturally expect, $\Sigma_*$ depends on $q_0$ almost only 
through $g(q_0)$:
 \begin{equation}
  \left.\Sigma_*(q)\right|_{\mbox{small $u$}}\approx
  -\frac{g(q_0)}{u\kappa}\ln\left(
  \frac{q_0-\varepsilon_h+i\delta}{q_0-\varepsilon_l+i\delta}
  \right).
 \end{equation}
  This particular expression uses the approximation that $g(E)$ is 
almost constant.
  $\varepsilon_{h,l}$ are the upper and lower limits of the electron 
dispersion relation, and are expected to be very far away from $q_0$.

  This finding, that $\Sigma_*$, and $\Sigma$ by a similar calculation, 
are almost independent of $q$, is quite intuitive. A slow magnon field 
tends to stay fixed at a point in space-time, and therefore its 
interaction is independent of $\mathbf{q}$. When $\Sigma$ is independent 
of $\mathbf{q}$, there is no divergence at thresholds (which depend on 
$\mathbf{q}$) and, equivalently, no singularities at thresholds in 
$q_0$.

  The divergent integral of eqn.~(\ref{eq:electron_self_energy_slow}) 
can therefore be treated as a constant to a good approximation. This 
gives rise to the renormalization of $\varepsilon$. $\Delta E$ is 
unaffected, and therefore there are no other corrections to $\Sigma$ at 
any order (consider the Dyson--Schwinger equations).

  The above argument is for the limiting case of zero spin-wave 
velocity. This is also intuitive from considering the BCS theory, where 
the electron-pair interaction is governed both by the coupling constant 
and the cut-off at the Debye frequency $\omega_D$ which corresponds to 
$u\kappa$ in the above case. There is no superconductivity in the limit 
of zero $\omega_D$. What, then, will be the case when $u\kappa$ is small 
but non-zero? We shall argue in the next section that even in that case, 
we cannot have electron-pair formation due to magnon exchange. Therefore 
electrons remain more or less free.
  Note that the formation of an electron-pair condensate should reflect 
in the electron Green's function through its imaginary part (and 
possibly also the real part).

  Thus, at least for the case of linear spin-wave dispersion relation, 
the behaviour of the electrons changes qualitatively depending on 
whether the spin-wave velocity is fast or slow.
  For fast magnons ($u\gg v_F$), the response of the (magnetic) vacuum 
is fast and collective, and the electrons are under its influence.
  For slow magnons ($u\ll v_F$), the response of the vacuum is slow and 
partial, and the electrons move around more freely.

  These results are quite intuitive, because according to 
eqn.~(\ref{eq_linear_spin_wave_velocity}), $u$ is large when $S$ is 
large and $g(\mu)$ is small.
  If the electron dispersion relation goes as $\epsilon\sim k^2/2m$, one 
can see that the $u>v_F$ when $v_F$ is small and $S$ is large. In other 
words, the case of fast spin-wave velocity occurs only near the Mott 
insulator state or some such state with large $S$ and small electron 
velocity.
  Spin-wave velocity is slow when $S$ is small, i.e., when the magnetic 
order is broken, so that the electrons move around more freely.

 \section{A study of high-$T_C$ superconductivity}
 \label{sec_highTc}

  Since high-$T_C$ cuprates are to a good approximation 2-d systems with 
AF interaction, let us focus on the case corresponding to these 
conditions.

  We saw at the end of the previous section how electrons start to 
become more free when the magnetic order is weakened.

  We then consider how superconducting pairs may be formed.

  In the exchange of a magnon between two electrons, the interaction is 
attractive between $e$ and $e_*$, but is repulsive between $e$ and $e$, 
or $e_*$ and $e_*$.
  This implies that for the formation of an $ee_*$ pair, the binding 
energy must exceed $\Delta E$, as otherwise there is insufficient energy 
to excite $e$ to $e_*$.
  But it would be unnatural that the interaction strength is so large as 
to allow such pairs to form.
  For instance, the BCS binding energy is expected to be less than the 
cut-off energy (the Debye frequency in BCS theory) which is given by 
$\Delta E$.
  If such an interaction arises at all, it can only be due to something 
like overscreening, and this circumstance is more naturally associated 
with SDW (spin-density wave) or some such magnetic structure.

  We thus believe that superconductivity through magnetic exchange 
occurs through a different mechanism, and this is through, let us say, 
the mixing between the two (e.g., spin-up and spin-down) sub-lattices 
that constitute the AF ground state (but note that a completely 
analogous argument can be made for ferromagnetism: this argument is not 
specific to AF).

  Through this mechanism, it is possible to `borrow' an $e_*$ electron 
from a neighbouring site to form spin-singlet pairs without paying the 
excitation energy $\Delta E$.

  Let us estimate the size of the superconducting gap due to this 
mechanism, using the BCS expression quoted in the introduction 
(eqn.~(\ref{eqn_electron_magnon_delta_naive})):
 \begin{equation}
  \Delta=2\omega_\mathrm{cut}\exp\left(-1/\mathcal{V N}(\mu)\right),
 \end{equation}
  where $\mathcal{V}$ is the effective 4-point interaction strength, and 
is estimated by $f^{-2}$ times some function $\Theta(\theta)$ of a 
`mixing-angle' for the `mixing' between the borrowed electrons and the 
initial electrons. The cut-off, $\omega_\mathrm{cut}$, is estimated by 
$\Delta E$. $\mathcal N(\mu)$ here is the number density of these 
`borrowed' electrons.
  Altogether, we thus expect
 \begin{equation}
  \Delta\sim\Delta E\exp
  \left(-g(\mu)/\Theta(\theta)\mathcal{N}(\mu)\right).
  \label{eq:BCSestimate}
 \end{equation}
  The expression which we shall obtain later on corresponds to 
$\mathcal{N}(\mu)=M_h/2\pi\hbar^2$, up to a possible constant factor.
  This is the number density of Higgs bosons if the chemical potential 
were exactly at the excitation energy of a $\mathbf{k}=0$ Higgs boson.

  There is one conceptual problem with this approach, namely that this 
mixing between the sub-lattices is not an electronic excitation but 
essentially an alteration of the AF ground state, in the sense that the 
flipping of the spin due to this effect is associated with the local 
modification of the vacuum from, say, $\Phi=(0,v)$ to $\Phi=(v,0)$, 
where $\Phi$ is an order parameter.
  In other words, the phenomenon that we are describing here is, we 
believe, Goldstone-boson condensation.
  What we need to discuss then is the stability of the AF ground state 
after Goldstone-boson condensation has taken place.
  This instability appears via the Higgs--Goldstone coupling which is 
due to the mixing of the Goldstone boson with the Higgs boson.
  This mixing is, as we shall see, a direct consequence of the 
Goldstone-boson condensation.

  A physical picture for this effect is as follows.
  In the AF state, moving an electron to its neighbouring site, without 
disturbing the AF order, requires flipping the spin of the electron.
  This takes place via the emission of a magnon.
  However, moving the electron without raising it to the spin-excited 
state corresponds to the Higgs boson, and so the Higgs boson couples to 
the Goldstone boson, or there is effectively a mixing between the Higgs 
boson and the magnon.

  We know how one might go about calculating this effect, but the full 
description is cumbersome, and we shall parametrize it by an effective 
mixing angle $\theta$.
  This allows us to calculate the corrections to the Higgs boson Green's 
function.

 \subsection{Higgs--Goldstone coupling}

  In our picture, Goldstone-boson condensation by itself costs energy, 
which is compensated by the superconducting condensation energy.
  $\phi$ condensation can at least formally be parametrized by a 
negative $\phi$ mass squared $-\mu_\phi$. This leads to the mass matrix:
  \begin{equation}
   M^2_{\phi,h}u^4=\left(\begin{array}{cc}
   -\mu_\phi & 0 \\ 0 & M_h^2u^4-\mu_\phi 
   \end{array}\right).
  \end{equation}

  But a Goldstone boson is necessarily massless, and so the physical 
modes are obtained by rotating $\phi$ and $h$ by a mixing angle 
$\theta_{h\phi}$ such that $\phi$ becomes massless.
  Some off-diagonal counter-terms are obviously necessary to stabilize 
the vacuum in this way, but let us skip the details here.

  Let us denote this angle $\theta_{h\phi}$ simply as $\theta$ in the 
following. $\theta$ is determined by the requirement that the total 
energy is minimized. That is, if superconductivity occurs, $\theta$ is 
such that the condensation energy plus the extra energy due to exciting 
the Goldstone bosons is minimized.
  This is clearly too complicated, and so let us treat $\theta$ as a 
free parameter in the following.

  We denote the mixed states $h^{'}$ and $\phi^{'}$ as
 \begin{equation}
  \left(\begin{array}{c}h^{'}\\i\phi^{'}\end{array}\right)=
  \left(\begin{array}{cc}\cos\theta&\sin\theta\\
  -\sin\theta&\cos\theta\end{array}\right)
  \left(\begin{array}{c}h\\i\phi\end{array}\right).
  \label{eqn_mixing_angle_definition}
 \end{equation}
  In general, only one linear combination of Goldstone bosons is 
involved, and the other Goldstone boson mode remains orthogonal to both 
$h^{'}$ and $\phi^{'}$.

  In terms of $\mu_\phi$ introduced above, we have
 \begin{equation}
  \tan\theta=\sqrt{\frac{\mu_\phi}{M_h^2u^4-\mu_\phi}}
  \approx\sqrt{\mu_\phi}/M_hu^2.
 \end{equation}
  Let us assume that $\mu$ is small and the mixing angle is not huge, so 
that $M_{h'}u^2$ can be approximated by $M_hu^2=\Delta E$ as calculated 
in sec.~\ref{subsec_higgs_boson}.
  In principle, $\theta$ can be measured by measuring the discrepancy 
between electronic excitation energy $\Delta E$ and the Higgs-boson 
excitation energy, but this is probably too delicate to be useful.

  In terms of $h$ and $\phi$, the $h\phi\phi$ coupling is given by
 \begin{equation}
  \frac{\hbar^2}4fh(f^{-1}\partial_\mu\phi_2)^2.
  \label{eqn_three_point_current_state}
 \end{equation}
  We derived this from the kinetic-energy part of the effective 
Lagrangian density for the order parameter field,
 \begin{equation}
  \mathcal{L}_\mathrm{kinetic}=
  \frac{\hbar^2}2(\partial_\mu\Phi_i)^2=
  \frac{\hbar^2}2\left[
  \left(\frac{\partial\Phi_i}{\partial t}\right)^2
  -u^2\left(\nabla\Phi_i\right)^2\right],
  \label{eqn_phi_kinetic_energy}
 \end{equation}
  using $v=f/2$ as before. Since the kinetic-energy part of the 
effective Lagrangian is largely model independent, we believe that this 
result is fairly general.

  We are interested in the $\phi^{'}h^{'}h^{'}$ coupling. From 
eqn.~(\ref{eqn_three_point_current_state}), for near on-shell Higgs 
bosons, and in terms of $h^{'}$ and $\phi^{'}$, the three-point coupling 
$\Gamma$ is found to be given by
 \begin{equation}
  \frac{\Gamma}{2!}\approx\frac{\hbar^2f^{-1}
  (M_{h^{'}}u^2)^2}4\sin\theta(1-3\cos^2\theta).
  \label{eqn_higgs_two_magnon_vertex_final}
 \end{equation}
  The $2!$ is the identical particle factor.
  Note that this $\Gamma$ is different from the electron-magnon vertex 
which appeared in sec.~\ref{sec_spin_current}.
  Even though there is, in principle, a four-point function in addition 
to the three-point interaction, we do not know how to sum both 
three-point and four-point vertices to all orders, and therefore we 
neglect it here.

  One simple estimate for $\theta$ is obtained by saying that $\Gamma^2$ 
is maximized.
  We find that $\Gamma^2$ is maximized when $\theta=\pi/2$, which is too 
large to be acceptable. A local maximum occurs for 
$\sin\theta=\sqrt{2}/3$, when $\theta$ is 28 degrees and 
$\sin^2\theta(1-3\cos^2\theta)^2=32/81$.
  This seems more appropriate, but let us not adopt any specific value 
here.

  For the sake of later analysis, let us introduce the dimensionless 
coupling $x$, defined by
 \begin{equation}
  x=\frac{-\Gamma^2}{4\pi M_{h'}^3u^6}\times\frac1{(\hbar u)^2}.
  \label{eqn_x_definition}
 \end{equation}
  Note again that this $x$ is different from the normalized energy which 
appeared in sec.~\ref{subsec_gribov_solution}.
  This is then given by
 \begin{equation}
  x\approx-\frac{f^{-2}M_{h'}u^2}{16\pi (\hbar u)^2}
  \sin^2\theta(1-3\cos^2\theta)^2
  =-\frac{\Delta E}{16\pi(\hbar u)^2g(\mu)}
  \sin^2\theta(1-3\cos^2\theta)^2.
  \label{eqn_x_estimation}
 \end{equation}
  We found in sec.~\ref{subsec_delta_e_S} that $\Delta E=S/g(\mu)a^2$ 
and that $\hbar u\pi/a=\Delta E$.
  We can use these relations to eliminate $\Delta E$ and $g(\mu)$:
 \begin{equation}
  x\approx -\frac{\pi}{16S}\sin^2\theta(1-3\cos^2\theta)^2.
  =-\mathcal{O}(10^{-2}\sim10^{-1})\frac1S.
 \end{equation}

  This is somewhat surprising in the sense that it implies that the 
dimensionless coupling $x$ grows as the magnetization falls. However, 
this behaviour is correct, as we shall show later, since this leads to 
the survival of magnetic order in the large-magnetiza\-tion region.

 \subsection{One-loop results}
 \label{sec_oneloop}

   Our objective is to analyze vacuum instability due to the exchange of 
Goldstone magnons.
  In two spatial dimensions and in the leading-order approximation, the 
potential due to the single exchange of a magnon is proportional to 
$\log(r)$ in the real space.
  The potential is therefore unbound as $r$ goes to infinity. This makes 
Higgs bosons long-distance, or infra-red, confined.
  This affects the way that the Higgs bosons can be excited.
  Presumably this disfavours single Higgs-boson excitations.

  This confinement only affects the excitations and does not affect the 
structure of the vacuum.
  Hence it is quite distinct from the superconductivity phenomenon which 
we are interested in, which is something that actually alters the 
structure of the vacuum.
  In short, we are interested in Higgs super-criticality, which occurs 
when the binding energy exceeds the energy that is needed to create a 
Higgs boson, and is an $\mathcal{O}(\hbar/M_h u)=\mathcal{O}(a)$ 
short-distance effect.
  We should add that the super-conducting coherence length is indeed 
relatively short in the real high-$T_C$ systems \cite{highTc}, which 
adds support to our discussion.

\begin{figure}[ht]{
  \centerline{
   \includegraphics[width=5cm]{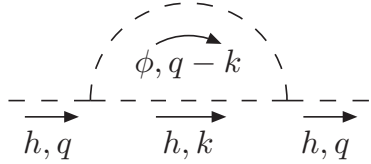}
  }
  \caption{The one-loop self-energy diagram.
  \label{fig_one_loop_self_energy}}}
 \end{figure}

  The finite ultra-violet contributions can be summed to all orders.
  Let us start with the one-loop self-energy correction to the Higgs 
boson propagator. The corresponding Feynman diagram is shown in 
fig.~\ref{fig_one_loop_self_energy}. All momenta are $2+1$-vectors, 
i.e., with two space and one time components.

  In the following, let us adopt the convention $\hbar=u=1$.
  It is easy to reintroduce $\hbar$ and $u$ by counting the dimensions.
  The square of a vector 2+1-vector is given by 
$q^2\equiv q_0^2-u^2\mathbf{q}^2\equiv q_0^2-\mathbf{q}^2$.
  The one-loop self-energy $\Sigma_1(q)$ is given by
 \begin{equation}
  \Sigma_1(q)=\Lambda_0\int\frac{d^3k}{(2\pi)^3i}\frac1{(q-k)^2+i\delta}
  \frac1{k^2-M_0^2+i\delta}.
  \label{eqn_one_loop_sigma_amplitude}
 \end{equation}
  $M_0$ is the bare mass of the Higgs boson, and is the same quantity as 
that denoted as $M_{h'}$ in the previous section. The change in the 
notation is so as to be more clear when discussing the mass 
renormalization.
  In the same way, let us call the bare coupling $\Gamma_0$. This is the 
same quantity as that denoted as $\Gamma$ in the previous subsection. 
  $\Lambda_0$ in the above equation is given by
 \begin{equation}
  \Lambda_0=-\Gamma_0^2.
  \label{eqn_lambda_convention}
 \end{equation}
  This is the bare coupling, to which we shall calculate the corrections 
which arise due to the vertex correction diagrams (and not due to the 
vacuum polarization diagrams).
  We have included the factor $2$ for the two Goldstone modes.
  As explained below eqn.~(\ref{eqn_higgs_two_magnon_vertex_final}), the 
factor $2$ for identical Higgs bosons has already been taken into 
account.
  The choice of the sign is such that positive $\Lambda_0$ corresponds 
to repulsion and negative $\Lambda_0$ to attraction.
  $\Lambda_0$ has the dimension $($energy$)^5\times$area.

  Let us say that $\mathbf{q}$ is small compared with the zone 
boundaries, so that we will not need to worry about the boundary 
effects.
  In this regard, $u$ is small compared with the electron velocity, but 
it is not small here since we are not comparing it against the electron 
velocity.
  Since $\hbar u\pi/a=\Delta E$, we do not have to worry about boundary 
effects so long as the Goldstone boson does not carry energy that is 
comparable with $\Delta E$.

  Eqn.~(\ref{eqn_one_loop_sigma_amplitude}) can be integrated easily 
using standard methods \cite{bjorkendrell}. First, the usual procedure 
of combining the denominators yields
 \begin{equation}
  \Sigma_1(q)=\Lambda_0
  \int_0^1d\lambda \int\frac{d^3\ell}{(2\pi)^3i}\left[\ell^2
  +\lambda\left(q^2(1-\lambda)-M_0^2\right)+i\delta\right]^{-2}.
 \end{equation}
  The Wick rotation yields
 \begin{equation}
  \Sigma_1(q)=\Lambda_0\int_0^1d\lambda
  \int\frac{d^3\tilde\ell}{(2\pi)^3}\left[-\tilde\ell^2
  +\lambda\left(q^2(1-\lambda)-M_0^2\right)+i\delta\right]^{-2}.
 \end{equation}
  After evaluating this Euclidean integral, we obtain
 \begin{equation}
  \Sigma_1(q)=\frac{\Lambda_0}{8\pi}\int_0^1d\lambda
  \left[\lambda\left(M_0^2-q^2(1-\lambda)\right)-i\delta\right]^{-1/2}.
 \end{equation}
  Finally, under the condition $M_0^2>q^2>0$, we obtain
 \begin{equation}
  \Sigma_1(q)=\frac{\Lambda_0}{4\pi\left|q\right|}
  \tanh^{-1}\frac{\left|q\right|}{M_0},
  \label{eqn_one_loop_sigma_result}
 \end{equation}
  where $\left|q\right|=\sqrt{q^2}$. Near $\left|q\right|=M_0$, this 
diverges logarithmically, reflecting the long-distance confinement.
  In the region $\left|q\right|< M_0$, $\Sigma$ has the following 
expansion:
 \begin{equation}
  \Sigma_1\approx\frac{\Lambda_0}{4\pi M_0^3}
  \left(M_0^2+\frac{q^2}{3}+\frac{q^4}{5M_0^2}+\cdots\right).
  \label{eqn_one_loop_sigma_series}
 \end{equation}
  Let us retain the first two terms. This approximation is accurate to 
$10\%$ up to $\left|q\right|/M_0=0.7$ or so.

  Since we have the following general relation:
 \begin{equation}
  G^{-1}=Z^{-1}(q^2-M^2)=q^2-M_0^2-\Sigma,
  \label{eqn_greens_function_general}
 \end{equation}
  we can immediately write down the one-loop $Z$ and $M^2$ as
 \begin{equation}
  Z^{-1}_1=1-\frac{\Lambda_0}{12\pi M_0^3},\qquad
  \frac{M^2_1}{M_0^2}=Z_1\left(1+\frac{\Lambda_0}{4\pi M_0^3}\right).
  \label{eqn_one_loop_Z_and_y}
 \end{equation}
  This corresponds to a (short-distance) correction to the Higgs-boson 
mass of the form
 \begin{equation}
  M_1^2=M_0^2-\Sigma_1\approx M_0^2+\frac{\Lambda_0}{4\pi M_0}
  =M_0^2(1+x).
 \end{equation}
  The dimensionless coupling $x$ is defined in 
eqn.~(\ref{eqn_x_definition}).

  For the sake of completeness, it is relatively easy to extend the 
preceding analysis to the case with an artificial small mass $m$ for the 
(pseudo-)Goldstone boson, which screens the interaction with a 
characteristic distance $\hbar/mu$. In this case, the one-loop 
self-energy correction is given by
 \begin{equation}
  \Sigma_1=\frac{\Lambda_0}{4\pi\left|q\right|}
  \tanh^{-1}\frac{\left|q\right|}{M_0+m}
  \label{eqn_one_loop_sigma_with_cutoff}
 \end{equation}
  We again see that the self-energy diverges near the threshold, 
corresponding to the long-distance correction mediated by the Goldstone 
boson which goes almost on shell.

\begin{figure}[ht]{
  \centerline{
   \includegraphics[width=5cm]{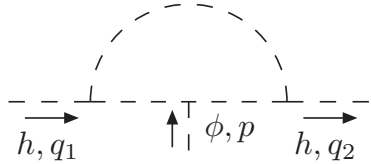}
  }
  \caption{The one-loop vertex-correction diagram.
  \label{fig_one_loop_vertex_correction}}}
 \end{figure}

  Let us now consider the vertex correction. The corresponding Feynman 
diagram is shown in fig.~\ref{fig_one_loop_vertex_correction}.

  The one-loop amplitude is given by
 \begin{equation}
  \Gamma_1/\Gamma_0-1=\Lambda_0\int\frac{d^3k}{(2\pi)^3i}
  \frac1{k^2+i\delta}\frac1{(q_1-k)^2-M_0^2+i\delta}
  \frac1{(q_2-k)^2-M_0^2+i\delta}.
  \label{eqn_one_loop_vertex_correction_amplitude}
 \end{equation}
  $\Gamma_1$ is $\Gamma_0$ plus the the one-loop correction.
  Proceeding in a similar fashion to that for the self-energy, we obtain 
the following Feynman parameter integral:
 \begin{multline}
  \Gamma_1/\Gamma_0-1=\frac{\Lambda_0}{16\pi}
  \int_0^1d\lambda_1d\lambda_2d\lambda_3
  \delta(1-\sum_{i=1}^3\lambda_i)\\
  \bigl[\left(\lambda_1M_0^2-q_1^2\lambda_2\lambda_3\right)+
  \left(\lambda_2M_0^2-q_2^2\lambda_1\lambda_3\right)
  -p^2\lambda_1\lambda_2-i\delta\bigr]^{-3/2}.
  \label{eqn_one_loop_vertex_correction}
 \end{multline}

  In the general case, eqn.~(\ref{eqn_one_loop_vertex_correction}) is 
difficult to handle analytically, even though the first two integrations 
are relatively easy. We have been able to evaluate it with the aid of 
the on-line software \cite{integrator} based on \textsc{Mathematica}, 
but the result is cumbersome and so we do not reproduce it here.

  Here we are only interested in the constant term which arises from 
eqn.~(\ref{eqn_one_loop_vertex_correction}). This corresponds to the 
limit in which the external momenta are small compared with $M_0$. In 
particular, when $p$ is small, the vertex-correction amplitude is 
related to the self-energy amplitude by
 \begin{equation}
  \Gamma_1/\Gamma_0-1=\frac{\partial\Sigma_1}{\partial M_0^2}.
  \label{eqn_vertex_self_energy_relation}
 \end{equation}
  This can be deduced from the form of 
eqn.~(\ref{eqn_one_loop_sigma_amplitude}).
  We then take $q^2\ll M_0$. The application of 
eqn.~(\ref{eqn_vertex_self_energy_relation}) to the leading term of 
eqn.~(\ref{eqn_one_loop_sigma_series}) yields
 \begin{equation}
  \Gamma_1/\Gamma_0-1\approx-\frac{\Lambda_0}{8\pi M_0^3}.
  \label{eqn_one_loop_vertex_correction_approx}
 \end{equation}

  From eqn.~(\ref{eqn_one_loop_Z_and_y}), the one-loop estimate for the 
critical coupling is $-\Lambda_0\approx4\pi M_0^3$, at which point the 
one-loop mass becomes zero, and above which the mass becomes tachyonic. 
  At this value of $-\Lambda_0$, by 
eqn.~(\ref{eqn_one_loop_vertex_correction_approx}), the one-loop vertex 
correction makes the effective vertex greater by $50\%$. Clearly, 
finite-order calculations are insufficient to deal with such corrections 
and so a non-perturbative method is needed.
  In the next subsection, we shall sum such contributions to all orders 
using the Dyson--Schwinger equations.

  Before proceeding, we note that a tachyonic Higgs mass by itself only 
suggests the instability of the AF ground state, and not the form of the 
true ground state. For instance, one may very well conclude at this 
stage that the true ground state is given simply by $v=0$, i.e., zero 
magnetization, which would be a reasonable interpretation for tachyonic 
Higgs mass. We shall show, however, that this is only partially true. In 
the all-order analysis, the real part of the Higgs-mass squared does 
indeed go negative above a certain value of the coupling, but 
ground-state instability occurs regardless of the sign of the real part 
of the Higgs-mass squared.

 \subsection{Dyson--Schwinger equations}
 \label{sec_dysonschwinger}

  We have shown that finite-order treatment is insufficient.
  In particular, we would like to sum the finite short-distance 
contributions to all orders.
  Let us deal with this problem by making use of the Dyson--Schwinger 
equations.

\begin{figure}[ht]{
  \centerline{
   \includegraphics[width=5cm]{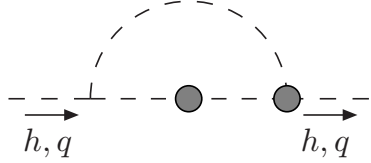}
  }
  \caption{The self-energy diagram corresponding to the Dyson--Schwinger 
equation. The blobs represent all-order corrections.
  \label{fig_dyson_schwinger_self_energy}}}
 \end{figure}

  First, the self-energy diagram is shown in 
fig.~\ref{fig_dyson_schwinger_self_energy}. This includes all possible 
perturbative corrections that involve the $hh\phi$ coupling, other than 
vacuum polarization which we either discard, or assume that it is taken 
into account in the Goldstone boson propagator.
  The non-perturbative equation is thus a modification of 
eqn.~(\ref{eqn_one_loop_sigma_series}), given by
 \begin{equation}
  \Sigma=(q^2-M_0^2)-Z^{-1}(q^2-M^2)=
  \frac{\Gamma\Gamma_0}{4\pi M^3}\left(M^2+\frac{q^2}{3}+\cdots\right).
  \label{eqn_dyson_schwinger_sigma}
 \end{equation}
  $M$, $\Sigma$ and $\Gamma=\sqrt{-\Lambda}$ are now all-order 
quantities.
  Eqn.~(\ref{eqn_dyson_schwinger_sigma}) sums only the finite terms in 
the higher-order corrections, and not the singular infra-red 
contributions.

  Let us introduce the following notation:
 \begin{equation}
 % x=\frac{\Gamma_0^2}{4\pi M_0^3},\qquad
  y=\frac{M}{M_0},\qquad
  w=\frac{\Sigma(q=0)}{M_0^2},\qquad
  g=\frac{\Gamma}{\Gamma_0}.
  \label{eqn_definition_of_variables}
 \end{equation}
  $y$ and $w$ are different from the quantities that appeared in 
sec.~\ref{subsec_gribov_solution}.
  $y$ is the normalized mass, which we would like to obtain as a 
function of the normalized coupling $x$ defined by 
eqn.~(\ref{eqn_x_definition}). $w$ is the normalized self-energy and 
$g$ gives the coupling renormalization.
  From eqn.~(\ref{eqn_dyson_schwinger_sigma}), we derive the following:
 \begin{equation}
  w=y^2Z^{-1}-1=\frac34(y^2-1),\qquad
  xyg=w(w+1).
  \label{eqn_w_relations}
 \end{equation}

\begin{figure}[ht]{
  \centerline{
   \includegraphics[width=10cm]{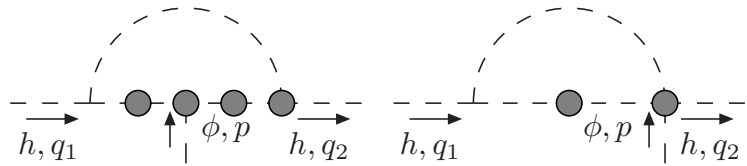}
  }
  \caption{The two diagrams whose sum contains all of the 
vertex-correction contributions.
  \label{fig_dyson_schwinger_vertex_correction}}}
 \end{figure}

  As for the vertex correction, we can write it as the sum of two 
diagrams, as shown in fig.~\ref{fig_dyson_schwinger_vertex_correction}. 

  The second of these diagrams contains a $4$-point vertex, and it would 
ordinarily be necessary to construct another Dyson--Schwinger equation 
to describe it. However, as we saw in sec.~\ref{sec_oneloop}, so long as 
we are only interested in the constant part of the corrections, the 
insertion of a Goldstone-boson vertex is formally equivalent to the 
derivative by $M_0^2$ and multiplication by $\Gamma_0$. That is, the sum 
of the two diagrams in fig.~\ref{fig_dyson_schwinger_vertex_correction} 
is given simply by
 \begin{equation}
  \Gamma/\Gamma_0-1=\frac{\partial\Sigma}{\partial M_0^2}
  \label{eqn_vertex_self_energy_relation_all_order}
 \end{equation}
  The partial derivative means differentiating by $M_0^2$ while keeping 
$\Lambda_0$ constant.
  Now $\Sigma$ is a function of both $\Lambda_0$ and $M_0^2$, but by 
counting the dimensions, we see that $\Sigma$ must be of the form
 \begin{equation}
  \Sigma=M_0^2w(x).
 \end{equation}
  Thus eqn.~(\ref{eqn_vertex_self_energy_relation_all_order}) can be 
written as
 \begin{equation}
  g=1+w-\frac{3x}2\frac{dw}{dx}.
  \label{eqn_vertex_vs_self_energy}
 \end{equation}

  To proceed, we make use of eqn.~(\ref{eqn_w_relations}) to eliminate 
$g$ from eqn.~(\ref{eqn_vertex_vs_self_energy}). This 
yields the following ordinary, first-order and non-linear differential 
equation:
 \begin{equation}
  \frac{3x}2\frac{dw}{dx}=(1+w)\left(1-\frac{w}{xy}\right).
  \label{eqn_w_vs_x_DE}
 \end{equation}
  In order to obtain $w=(3/4)(y^2-1)$ as a function of $x$, we then need 
to solve eqn.~(\ref{eqn_w_vs_x_DE}) with the boundary condition
 \begin{equation}
  w(0)=0.
  \label{eqn_w_vs_x_boundary}
 \end{equation}
  Unfortunately, this boundary condition makes the solution of 
eqn.~(\ref{eqn_w_vs_x_DE}) difficult. We do not think that 
eqn.~(\ref{eqn_w_vs_x_DE}) can be integrated analytically and, on the 
other hand, numerical implementation suffers from instability near $x=0$ 
where eqn.~(\ref{eqn_w_vs_x_DE}) contains $0/0$.

  Let us adopt a power-series solution.
  Simply expanding $y$ in powers of $x$ and substituting in 
eqn.~(\ref{eqn_w_vs_x_DE}), we are able to read off the coefficients 
with the boundary condition given by eqn.~(\ref{eqn_w_vs_x_boundary}). 
  We used \textsc{Maxima} \cite{maxima} for preliminary algebra 
and the evaluation of the first 28 coefficients, and \textsc{Form} 
\cite{form} after that.
  Using \textsc{Form}, we calculated the first 102 coefficients, which, 
in our implementation, is equivalent to having verified the first 101. 
On a \textsc{Linux} machine equipped with a 2.6~GHz Pentium 4 processor, 
the evaluation of the last coefficient took just over 5 minutes to 
perform.
  The machine has 2~GB memory, but less than 1~\% of the memory was 
taken up by the calculation.
  The first few coefficients are given by
 \begin{equation}
  y=1+\frac23x-\frac79x^2+2x^3-\frac{641}{81}x^4+\frac{2321}{54}x^5
  -\mathcal{O}(x^6).
  \label{eqn_y_vs_x_power_series}
 \end{equation}
  It should be noted that $x$ being the normalized coupling, the 
expansion in $x$ should match with the finite terms from the 
perturbative calculation, order by order. For example, the truncation of 
eqn.~(\ref{eqn_y_vs_x_power_series}) at the first order in $x$ 
reproduces the one-loop perturbative result, 
eqn.~(\ref{eqn_one_loop_Z_and_y}), at the first power order in 
$\Lambda_0$.

  This series is divergent for all non-zero $x$. In other words, the 
critical coupling $x_c$, which is calculated as
 \begin{equation}
  x_c=\lim_{n\to\infty}\frac{a_{n-1}}{a_n},
  \label{eqn_radius_of_convergence}
 \end{equation}
  vanishes.
  Here, $a_n$ stands for the coefficient of the $n$'th power of $x$.
  The radius of convergence of the series is given by 
$\left|x_c\right|$.

\begin{figure}[ht]
  \centerline{\includegraphics[width=12cm]{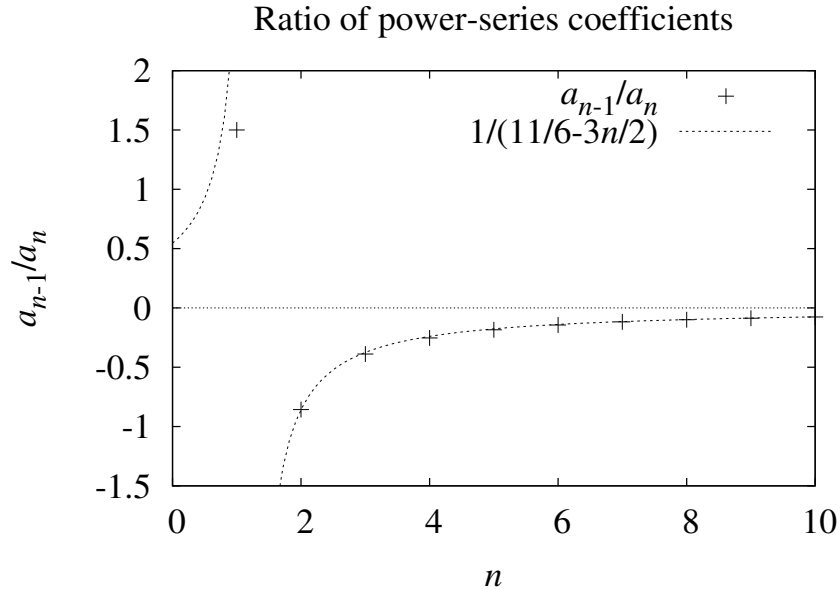}}
  \caption{The ratio $a_{n-1}/a_n$, for $1\le n\le10$, compared with the 
expression $(11/6-3n/2)^{-1}$.
  \label{fig_ratio_coeffs}}
 \end{figure}

  We plot the ratio appearing in eqn.~(\ref{eqn_radius_of_convergence}), 
for finite $n$, with $n$ up to $10$ in fig.~\ref{fig_ratio_coeffs}. We 
have found that these are numerically consistent with
 \begin{equation}
  \frac{a_{n-1}}{a_n}=\left[\frac{11}6-\frac{3n}2+
  \mathcal{O}(n^{-1})\right]^{-1}.
  \label{eqn_radius_of_convergence_behaviour}
 \end{equation}
  The coefficient of the $\mathcal{O}(n^{-1})$ contribution is about 
$0.3$, numerically.
  We see that for large $n$, the series is factorially divergent, and 
the radius of convergence is therefore zero.

 \subsection{Borel summation}
 \label{sec_borel_summation}

  A method which is commonly used to assign a meaning to factorially 
divergent expansions is that of Borel summation.
  For $Y(X)$ given by the power series
 \begin{equation}
  Y(X)=\sum_{n=0}^\infty A_nX^{-n},
 \end{equation}
  the Borel transformed series is given by
 \begin{equation}
  \mathcal{B}Y(t)=\sum_{n=0}^\infty\frac{A_{n+1}}{n!}t^n.
 \end{equation}
  $Y(X)$ is then recovered by the Laplace transform:
 \begin{equation}
  Y(X)=A_0+\int_0^\infty \mathcal{B}Y(t)e^{-Xt} dt.
  \label{eqn_laplace_transform}
 \end{equation}
  Since eqn.~(\ref{eqn_laplace_transform}) requires $X>0$, let us 
define
 \begin{equation}
  X=1/x,\quad A_n=a_n,\quad y(x)=Y(1/x),
 \end{equation}
  when $x>0$, and
 \begin{equation}
  X=-1/x,\quad A_n=(-1)^{n+1}a_n,\quad y(x)=-Y(-1/x),
 \end{equation}
  when $x<0$. 
  We can then readily find the Borel transformed series 
$\mathcal{B}Y(t)$.

\begin{figure}[ht]
  \centerline{\includegraphics[width=12cm]{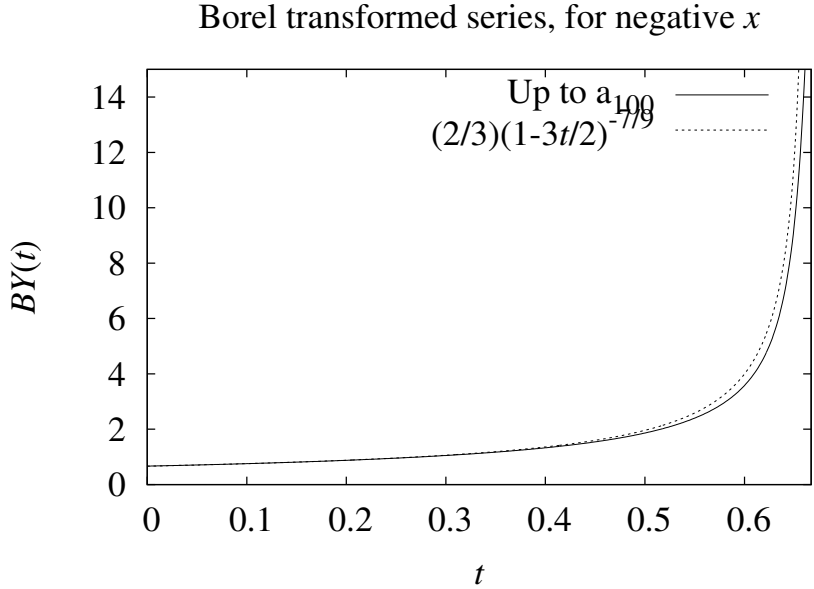}}
  \caption{$\mathcal{B}Y(t)$ for negative $x$, using the actual values 
of $a_n$ (solid line, with $n\le27$), and the estimation of 
eqns.~(\ref{eqn_borel_sum_estimate}) (dotted line).
  \label{fig_borel}}
 \end{figure}

  Alternatively, using the first two terms of 
eqn.~(\ref{eqn_radius_of_convergence_behaviour}), we can estimate the 
Borel sum to be
 \begin{equation}
  \left\{
  \begin{array}{l}
   \mathcal{B}Y(t)\approx a_1(1+3t/2)^{-7/9}\qquad (x>0),\\
   \mathcal{B}Y(t)\approx a_1(1-3t/2)^{-7/9}\qquad (x<0).
  \end{array}
  \right.
  \label{eqn_borel_sum_estimate}
 \end{equation}
  In fig.~\ref{fig_borel}, we compare this estimation, with $a_1=2/3$, 
against the series expansion using the actual values of $a_n$ up to 
$a_{100}$.
  For the range of $\mathcal{B}Y(t)$ shown in fig.~\ref{fig_borel}, the 
contribution of the last term in the expansion is at most about $0.5\%$ 
of the total number, and therefore we consider the error due to the 
truncation of the power series to be negligible.
  The two expressions match reasonably well.
  We shall henceforth make use of the estimation of 
eqns.~(\ref{eqn_borel_sum_estimate}) in our discussions.

  The first of eqns.~(\ref{eqn_borel_sum_estimate}), which corresponds 
to positive $x$, has a unique Laplace transform. In other words, the 
series is Borel summable. The result of the Laplace transform is
 \begin{equation}
  y(x)=a_0+a_1(2/3)^{7/9}e^{2/3x}x^{2/9}\Gamma(2/9,2/3x) \qquad (x>0),
  \label{eqn_laplace_transform_result_positive_x}
 \end{equation}
  again using the approximation of taking the first two terms of 
eqn.~(\ref{eqn_radius_of_convergence_behaviour}).
  $\Gamma$ is the upper incomplete Gamma function, defined by
 \begin{equation}
  \Gamma(s,a)=\int_a^\infty t^{s-1}e^{-t}dt.
  \label{eqn_incomplete_gamma_definition}
 \end{equation}

\begin{figure}[ht]
  \centerline{
   \includegraphics[width=4.5cm]{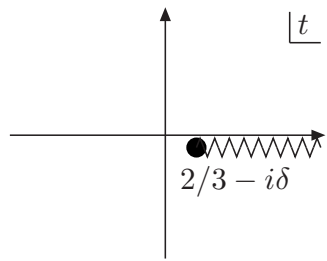}
  }
  \caption{Moving the branch cut off the real axis. The contour of 
integration is along the positive real axis.\label{fig_branchcut}}
 \end{figure}

  On the other hand, the second of eqns.~(\ref{eqn_borel_sum_estimate}), 
which corresponds to negative $x$, has a branch-point singularity at 
$t=2/3$. In other words, the series is non-Borel summable. We can still 
derive the expression for $y(x)$ formally and, not too surprisingly, the 
result turns out to be of the same form as 
eqn.~(\ref{eqn_laplace_transform_result_positive_x}), but is dependent 
on the choice of the contour of integration.
  Let us choose the contour such that we obtain a negative imaginary 
part for the self-energy, corresponding to the retarded Green's 
function. This choice of the contour is obtained by moving the 
singularity downwards in the complex $t$ plane, as shown in 
fig.~\ref{fig_branchcut}.

\begin{figure}[ht]
  \centerline{\includegraphics[width=12cm]{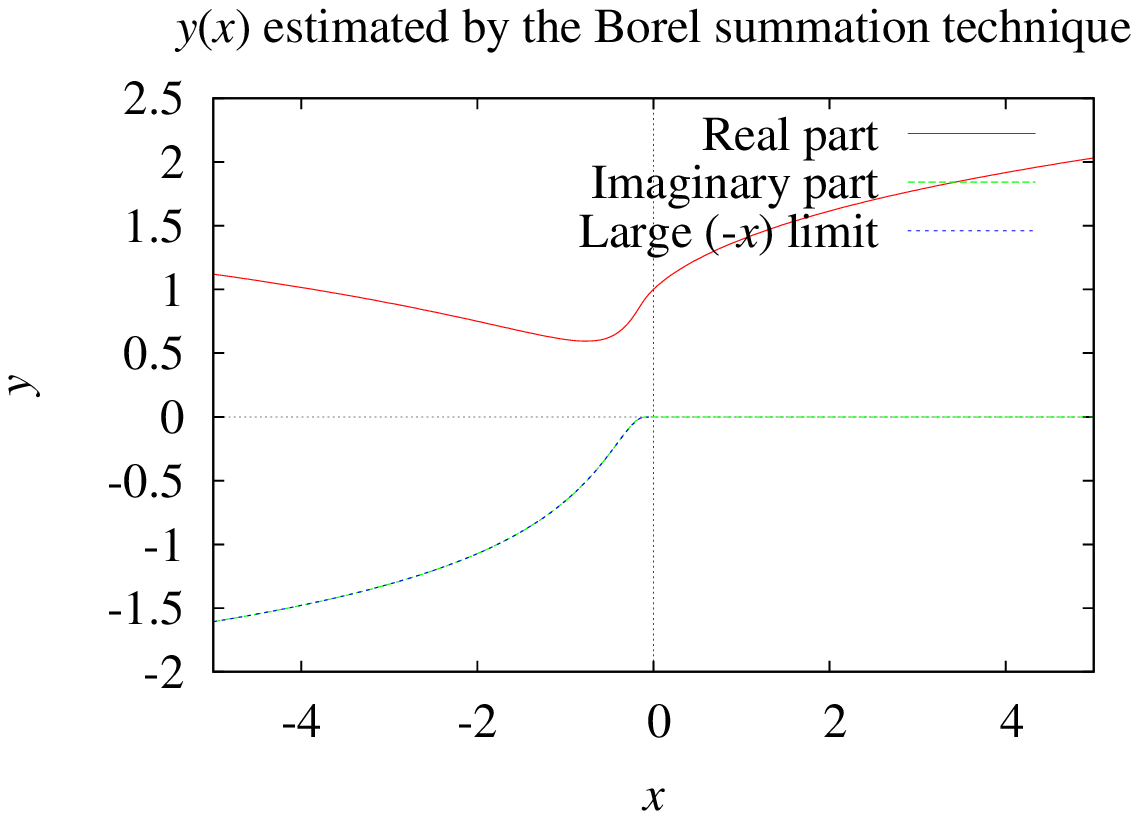}}
  \centerline{\includegraphics[width=12cm]{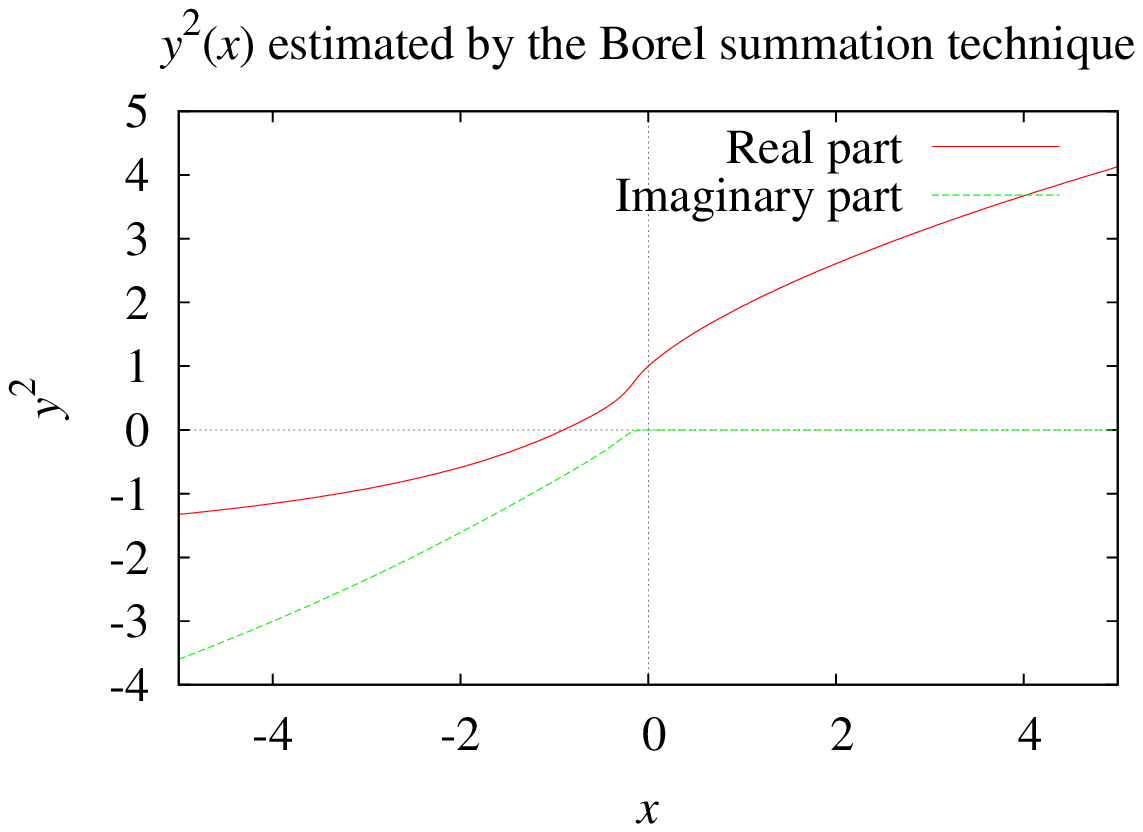}}
  \caption{$y(x)$ (upper figure), and $y^2(x)$ (lower figure), evaluated 
as the Laplace transform of eqn.~(\ref{eqn_borel_sum_estimate}).
  \label{fig_borelsum}}
 \end{figure}

  This is equivalent to giving a negative and infinitesimal imaginary 
part to $x$, and so in the end, we have, for both positive and negative 
$x$, as well as $x=0$,
 \begin{equation}
  y(x)=
  1+\frac49\exp\left(\frac{2/3}{x-i\delta}\right)
  \left(\frac{2/3}{x-i\delta}\right)^{-2/9}
  \Gamma\left(\frac29,\frac{2/3}{x-i\delta}\right).
  \label{eqn_laplace_transform_result_all_x}
 \end{equation}
  The definition of the incomplete Gamma function, as given by 
eqn.~(\ref{eqn_incomplete_gamma_definition}), is ambiguous for negative 
or complex $a$. A more general formula reads
 \begin{equation}
  \Gamma(s,z)=\Gamma(s)-\gamma(s,z)
  =\Gamma(s)-\sum_{k=0}^\infty\frac{(-1)^k}{k!}\frac{z^{s+k}}{s+k},
  \label{eqn_incomplete_gamma_evaluation}
 \end{equation}
  using which we are able to evaluate $y(x)$ numerically. $\gamma(s,z)$ 
is the lower incomplete Gamma function.
  The resulting form of $y(x)$, and $y^2(x)$, is shown in 
fig.~\ref{fig_borelsum}.

  In fig.~\ref{fig_borelsum}, we also show the function
 \begin{equation}
  -(4/9)\Gamma(2/9)\sin(2\pi/9)e^{2/3x}(-2/3x)^{-2/9},
  \label{eqn_BCS_behaviour}
 \end{equation}
  for negative $x$.
  This is an $x\to-\infty$ limit of the imaginary part of 
eqn.~(\ref{eqn_laplace_transform_result_all_x}), and it agrees well with 
the imaginary part of $y(x)$. In general, we expect that any singularity 
at $t=t_0$ in the Borel transform induces a term of the form 
$e^{t_0/x}$.
  There is nothing new in this divergent behaviour of the perturbation 
series, the so-called renormalon \cite{renormalon} being a celebrated 
example.
  As for the exponential form of the Borel sum, since the imaginary part 
of the Green's function corresponds to the binding energy of the 
condensate to which the AF vacuum decays, this implies a BCS-like 
behaviour of the energy gap, as given by eqn.~(\ref{eq:BCSestimate}).

  We note that, unlike in finite-order calculations, the criticality, in 
the sense of an imaginary part of $y$, develops already at $x=-0$, even 
though it is quite small until $-x$ becomes comparable with $2/3$.

  The real part of $y$ remains positive even at large $-x$. This is 
against our naive expectation that strong attractive interaction makes 
the single-particle mass negative. The behaviour seen here suggests that 
there is a balance between the attractive interaction mediated by the 
Goldstone boson and the contribution due to the condensate which 
counteracts it. The former is dominant at small $-x$, and the latter is 
dominant at large $-x$.

  The real part of $y^2$ becomes negative at $x\approx-0.9$. Naively 
speaking, AF order co-exists with the new condensate 
state for $x\gtrsim-0.9$, and is destroyed below this value of $x$.

 \subsection{Phenomenological discussions}
 \label{sec_discussion}

  In the preceding sections, we were able to deduce the non-perturbative 
behaviour of the single-particle Green's function for the Higgs boson. 

  As per the conventional wisdom \cite{agd}, the imaginary part of the 
self-energy determines the relaxation time of the unstable ground state 
which, in turn, and by the uncertainty principle, is the inverse of the 
binding energy.
  The energy gap is thus given simply by
 \begin{equation}
  2\Delta=-\mathrm{Im}\,M\equiv-M_0u^2\mathrm{Im}\,y.
  \label{eqn_energy_gap}
 \end{equation}

  The total condensation energy, per unit area, is given by the 
two-dimensional phase-space integral
 \begin{equation}
  \frac{E_\mathrm{cond}}{A} = - \int\frac{d^2\mathbf{k}}{(2\pi)^2}
  \mathrm{Im}\sqrt{\mathbf{k}^2+M^2}
  = - \frac1{6\pi}\mathrm{Im}\left((K^2+M^2)^{3/2}-M^3\right).
  \label{eqn_cond_energy}
 \end{equation}
  $K$ stands for the upper limit of integration, which is close to 
$\pi/a$.
  In the limit as the $K$ tends to infinity, we obtain
 \begin{equation}
  \frac{E_\mathrm{cond}}{A} = - \frac1{4\pi}K\mathrm{Im}\,M^2.
 \end{equation}
  This expression, for very large $-x$, has the behaviour $(-x)^{4/9}$, 
or proportional to $\Delta^2$.

  We do not know how to calculate $T_C$, since long-range order is known 
to disappear for finite temperature in two spatial dimensions.

  Let us define $S_0$ by
 \begin{equation}
  -2/3x=S/S_0.
 \end{equation}
  By eqn.~(\ref{eqn_x_estimation}), we have
 \begin{equation}
  S_0=\frac{3\pi}{32}\sin^2\theta(1-3\cos^2\theta)^2,
  \label{eqn_S_zero_evaluated}
 \end{equation}
  and so $S_0$ is $\mathcal{O}(10^{-2})$ to $\mathcal{O}(10^{-1})$.

\begin{figure}[ht]
 \centerline{\includegraphics[width=12cm]{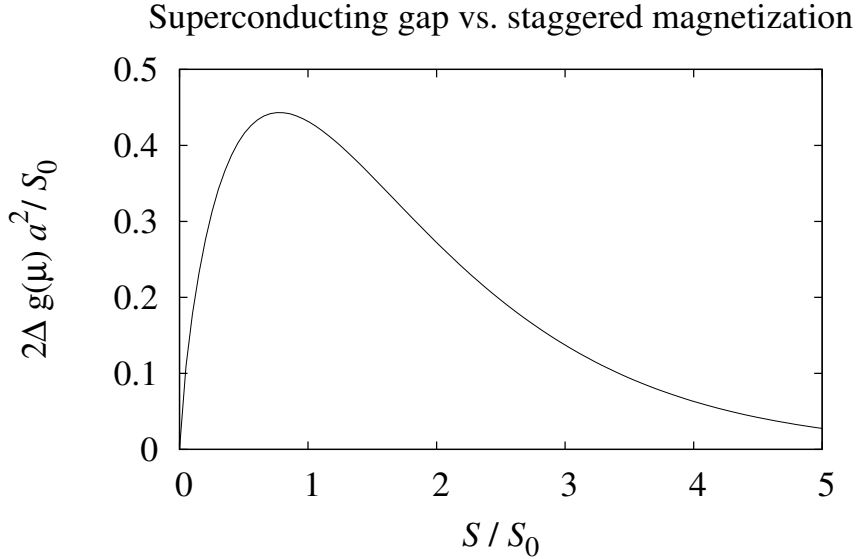}}
 \caption{The superconducting gap, normalized to $c_SJ^2a^2g_F$, as a 
function of the mean staggered magnetization $S$ normalized to 
$S_0$.\label{fig_Delta_S}}
 \end{figure}

  Using eqn.~(\ref{eqn_BCS_behaviour}) and then using 
eqn.~(\ref{eq:delta_e_S}), the superconducting gap is estimated as
 \begin{equation}
  2\Delta_\mathrm{SC}=1.17\Delta E e^{-S/S_0}(S/S_0)^{-2/9}
  =\frac{1.17S_0}{g(\mu)a^2} e^{-S/S_0}(S/S_0)^{7/9}.
  \label{eqn_Delta_S_numerical}
 \end{equation}
  This is shown in fig.~\ref{fig_Delta_S}.
  For large $S$, the exponential factor tends to suppress 
superconductivity, whereas for small $S$, the $\Delta E$ factor 
suppresses it. The maximum of $2\Delta_\mathrm{SC}$ as a function of 
$S/S_0$ occurs at $S/S_0=7/9$.

  Small $S$ and large $S$ correspond to over-doping and under-doping, 
respectively, and therefore the general behaviour is consistent with the 
doping behaviour of high-$T_C$ superconductors, assuming that $T_C$ is 
proportional to $\Delta$.
  Out of the parameters in eqns.~(\ref{eqn_S_zero_evaluated}, 
\ref{eqn_Delta_S_numerical}), $S_0$ can in principle be measured through 
the measurement of the magnetization at the peak of 
$2\Delta_\mathrm{SC}$, but one needs to be careful here since $S$ refers 
to the magnetization when superconductivity is not present. Perhaps some 
extrapolation of the magnetization at higher temperatures can be made to 
the magnetization at 0~K.
  This, together with the measurement of $g(\mu)$ and $a$, gives us 
$2\Delta_\mathrm{SC}$ as a function of $S$, to be compared against 
experiment.

  In the sense of the positivity of the real part of the renormalized 
$M^2=M_0^2y^2$, superconductivity co-exists with AF down to about 
$x=-0.90$, or $S/S_0=0.74$ ($0.74<7/9=0.77\cdots$).
  AF disappears in the over-doping region.
  The pseudo-gap is naturally associated with the excitation of $e$ to 
$e_*$, which is characterized by $\Delta E$.

  As for the overall magnitude of $\Delta_\mathrm{SC}$, an 
$\mathcal{O}(100~\mathrm{meV})$ superconducting gap is obtained when 
$1/a^2g(\mu)$ is $\mathcal{O}(1\sim10~\mathrm{eV})$. This seems quite 
reasonable. The superconducting gap goes up when the electron density 
goes down (against naive expectation). That is, smaller values of 
$1/m_\mathrm{eff}$ is favoured for the effective electron mass.

  As for the symmetry of the superconducting gap, our analysis does not 
single out a particular type of symmetry, but we expect, naively, that 
the requirement of forming spin-singlet pairs in AF background restricts 
pair formation in the $(\pm\pi,\pm\pi)$ directions, and so there are 
necessarily nodes in these directions.

 \section{Conclusions}
 \label{sec_conclusions}

  We made an adaptation of Gribov's axial-current conservation 
techniques to the situation of spin-current conservation in systems with 
partial magnetic order.
  We obtained the form factor, or the coupling strength of the Goldstone 
bosons, and the mass of the Higgs boson in a non-perturbative manner.
  We also obtained relationships among the parameters of the theory.
  Our analysis is of a general nature and rely only on conventional 
wisdom relating to quasiparticle dynamics in the momentum space, but 
many of the results were obtained using the approximation that $\Delta 
E$, the spin excitation energy, is independent of energy or momentum.
  Some other results depend on the assumption that the inverse of the 
electron Green's function is linear in energy.
  These assumptions are reasonable when the spin-wave velocity is small.

  Our approach is distinct from methods that start from Hamiltonians 
with localized states as the bases, because the Pauli exclusion 
principle acting on localized states is an ingredient of the latter 
formulations. The Pauli exclusion principle is present in our approach 
without such restrictions, in the form of the Feynman propagators, i.e., 
in the form of the analytical behaviour of the Green's functions.

  We then discussed two applications of our framework.

  We first discussed the case of fast magnons, i.e., near the Mott 
insulating state, with linear dispersion relation in three spatial 
dimensions. This case admits a treatment using the Gribov equation, 
which sums the leading overlapping divergences to all orders.
  The solution of the Gribov equations leads to a peculiar solution in 
which the fermionic degrees of freedom seem to disappear, and we are 
left with something which can be interpreted as a tower of bosonic 
excitations.
  A natural interpretation for this phenomenon is as a string-like 
one-dimensional excitation, which may correspond to the path 
transgressed by a wrong-spin electron.
  This phenomenon occurs when the spin-wave velocity is large, which 
implies large magnetization and small carrier density, i.e., near the 
Mott insulator state.

  With increased carrier density, the system starts to lose the magnetic 
order, and the spin-wave velocity starts to fall.
  When the spin-wave velocity falls below the electron velocity, i.e., 
$u<v_F$, the behaviour of the system changes qualitatively, because the 
collective response of the vacuum subsides and the electrons become more 
free, even though magnetization still remains. This cross-over occurs 
for reasonably large values of $S$.
  We have found that $u<v_F$ is satisfied unless $g(\mu)$ is small, 
$v_F$ is small and $S$ is large, i.e., quite near the Mott insulator 
state.
  This finding supports our being able to use the nearly-free electron 
picture to discuss high-$T_C$ superconductivity.

  When discussing high-$T_C$ superconductivity, we proposed that there 
is a mixing between the Goldstone and Higgs modes due to Goldstone-boson 
condensation.
  Goldstone-boson condensation is stabilized by superconductivity, and 
superconductivity cannot arise without Goldstone-boson condensation.
  This gives rise to an effective Higgs--Higgs--Goldstone three-point 
interaction.

  We then discussed the stability of the AF ground state with respect to 
the collapse due to this three-point interaction, which is attractive.
  In view of the observation of high-$T_C$ superconductivity in 
two-dimensional AF systems, we focused on the two-dimensional case, 
which is easier in the sense that the short-distance integral converges.

  The exchange of Goldstone magnons induces non-analytic\-ity in the 
Higgs-boson Green's function. This implies that the AF ground state is 
unstable at low temperature.
  The analysis of the Higgs-boson Green's function was carried out in an 
all-order manner, by using a Dyson--Schwinger summation of the relevant 
finite, short-distance, contribution to all orders.
  We found a solution for the mass renormalization as a power series in 
the coupling constant. The power series is factorially divergent, but 
can be treated using the Borel summation method.

  We found that an essential singularity develops in the mass 
renormalization when the coupling constant is zero.
  We thus conclude that the true AF ground state in two spatial 
dimensions contains a condensate into which the AF state decays.
  The energy gap associated with this condensate has the behaviour that 
is expected for superconductivity, and we argued that it indeed is the 
superconducting pair condensate.
  Superconductivity co-exists with AF in the under-doped region.

  A number of results of this paper can be tested experimentally.
  First, the expression for the superconducting energy gap.
  Second, the presence of the Higgs boson at larger values of carrier 
concentration, and the prediction of its mass.
  Third, the tower of bosonic states in AF systems at smaller values of 
carrier concentration, especially in the case of three spatial 
dimensions.
  In addition, the analytical relationships between such quantities as 
$\Delta E$, $g(\mu)$, $S$ and $u$ are all testable.

  What might be the signals of Goldstone-boson condensation? We can at 
the moment only think of indirect signatures such as the lowering of the 
Higgs-boson excitation energy from $\Delta E$, which may be too delicate 
to be treated as a concrete signal.

  Large $T_C$ can be achieved by making $g(\mu)$ small, or by reducing 
the effective mass of the electrons, without destroying the magnetic 
interaction.

 \begin{acknowledgments}

  We thank I.~Hase, M.~Hashimoto, S.~Sharma, K.~Yamaji, and all members 
of the Condensed Matter Physics group, AIST for extensive, informative 
and penetrating comments and discussions.

 \end{acknowledgments}

 \end{document}